\documentclass[preprint,12pt,doublespace]{revtex4}
\usepackage{rotating,booktabs}
\usepackage{natbib}
\usepackage{times,setspace}
\usepackage{amssymb,amsmath,graphicx}
\usepackage{graphicx}
\usepackage{amsmath,amssymb, amsthm}
\usepackage{rotating}
\usepackage{psfrag}
\usepackage[usenames]{color}
\usepackage{bm}

\newcommand{\bee}{\begin{equation}}
\newcommand{\ee}{\end{equation}}

\usepackage{hanging}

\begin{document}

\title{A universal mechanism generating clusters of differentiated loci during divergence-with-migration}
\author{Marina Rafajlovi{\'c}$^{1,2,*}$, Anna Emanuelsson$^{1}$, Kerstin Johannesson$^{3,2}$, Roger K. Butlin$^{4,2}$, and Bernhard Mehlig$^{1,2}$\\
$^1$\emph{\small Department of Physics, University of Gothenburg, SE-412 96 Gothenburg, Sweden}\\
$^2$\emph{\small The Linnaeus Centre for Marine Evolutionary Biology, University of Gothenburg, SE-405 30 Gothenburg, Sweden}\\
$^3$\emph{\small Department of Marine Sciences, University of Gothenburg, Tj\"arn\"o SE-452 96 Str\"omstad, Sweden}\\
$^4$\emph{\small Department of Animal and Plant Sciences, University of Sheffield, Sheffield S10 2TN, UK}\\
$^*$\emph{\small Corresponding author, {\tt marina.rafajlovic@physics.gu.se}}\\
\vspace*{1cm}
\small{\emph {Running head}: Origins of concentrated genetic architectures.}
}

\newpage
\begin{abstract}
Genome-wide patterns of genetic divergence reveal mechanisms of adaptation under gene flow. Empirical data show that divergence is mostly concentrated in narrow genomic regions. This pattern may arise because differentiated loci protect nearby mutations from gene flow, but recent theory suggests this mechanism is insufficient to explain the emergence of concentrated differentiation during biologically realistic timescales. Critically, earlier theory neglects an inevitable consequence of genetic drift: stochastic loss of local genomic divergence. Here we demonstrate that the rate of stochastic loss of weak local differentiation increases with recombination distance to a strongly diverged locus and, above a critical recombination distance, local loss is faster than local `gain' of new differentiation. 
Under high migration and weak selection this critical recombination distance is much smaller than the total recombination distance of the genomic region under selection. Consequently, divergence between populations increases by net gain of new differentiation within the critical recombination distance, resulting in tightly-linked clusters of divergence. The mechanism responsible is the balance between stochastic loss and gain of weak local differentiation, a mechanism acting universally throughout the genome. Our results will help to explain empirical observations and lead to novel predictions regarding changes in genomic architectures during adaptive divergence.

\end{abstract}

\newpage
\maketitle
\newpage
\section{Introduction}
In spite of substantial gene flow, populations under differential selection in a heterogeneous environment may diverge as partial barriers to gene exchange establish along the genome at loci involved in local adaptation (Barton and Bengtsson 1986). If the combined effects of these barriers are strong enough, gene flow may eventually cease and result in ecological speciation (Nosil 2012; Flaxman et al. 2013). Local adaptation of populations is observed everywhere in nature (Savolainen et al. 2013), but the genetic mechanisms involved at various stages of differentiation remain poorly understood. In particular, it is not known what mechanisms allow populations under differential selection and gene flow to diverge and, potentially, evolve into distinct species (Seehausen et al. 2014). This may depend, in part, on the genomic architecture of adaptive divergence (Smadja and Butlin 2011). Genome scans reveal that different species have very different numbers of loci that cause traits to diverge, ranging from one or a few loci of large effect to hundreds of loci each with presumably smaller effect  (Seehausen et al. 2014; Marques et al. 2016). A very intriguing empirical observation is that loci exhibiting divergence may not be uniformly distributed across the genome  (Via 2009; Feder et al. 2012; Feder et al. 2013; Seehausen et al. 2014). Instead `islands of divergence' or `clustered genetic architectures' are commonly observed (Feder et al. 2012; Jones et al. 2012; Feder et al. 2013; Marques et al. 2016), 
while there are few examples of divergent ecotypes in which observed genetic differentiation appears homogeneous (reviewed by Feder et al. 2013, but see Soria-Carrasco et al. 2014). Unveiling the mechanisms involved in establishing a non-uniform distribution of divergent loci is a key step towards understanding both local adaptation and speciation under gene flow. 

Gene flow due to migration between populations subject to divergent selection opposes differentiation. However, if divergence is established at one or a few loci, the effective migration rate in the genomic regions surrounding these loci is reduced due to linkage (Bengtsson 1985; Barton and Bengtsson 1986). For an illustration of this effect in infinitely large populations see, for example, Fig. 3b in Barton and Bengtsson (1986) and Fig. 1 in Feder and Nosil (2010) (note that Feder and Nosil (2010) simulated populations without drift, varying population size only to infer effects of gene flow on levels of differentiation). An instructive interpretation of these results is that the effect of indirect selection (the source of which is a diverged locus) weakens as the recombination distance from this locus increases.  By contrast to infinitely large populations,  this linkage to a diverged locus has two consequences in populations of finite size where random genetic drift is necessarily at work.  
First, the establishment probability of a new beneficial mutation is higher for the mutation landing closer to an already diverged locus than further away (hereafter, {\it the establishment bias}). For an illustration of this effect, see Fig.~3 in Feder et al. (2012).
Second, genetic drift results in stochastic loss of differentiation as the same allele may become fixed in both diverging populations. This effect may be opposed by linkage to another differentiated genomic region under divergent selection. As a consequence, the rate of  stochastic loss of differentiation may be larger at larger recombination distances from another diverged genomic region (see Aeschbacher and B\"urger (2014) for an analysis of this effect in a mainland-island model of divergence). Both the establishment-bias and the stochastic-loss effect necessarily influence the genetic patterns emerging during divergence-with-migration. Yet, earlier theoretical studies have focused only on understanding the importance of the establishment-bias effect (Yeaman and Whitlock 2011; Feder et al. 2012; Feder et al. 2013; Yeaman 2013). Disregarding stochastic loss has, for example, led Feder et al. (2012) to conclude that clustering of differentiated loci may occur only during early stages of divergence-with-migration because this is when the establishment bias is strong. These authors further conclude that, as divergence is ongoing, the establishment of new mutations becomes facilitated over the whole genome, and  the establishment bias inevitably weakens (see Figs. 5d-f, 6 in Feder et al. 2012). Consequently, these authors predict that clusters of differentiated loci  disappear during late stages of divergence and, instead, genome-wide, uniformly distributed differentiation appears  (Feder et al. 2012). This effect is referred to as {\it genome hitchhiking} by Feder et al. (2012). However, this genome-hitchhiking prediction is difficult to reconcile with earlier results of  multi-locus  simulations of divergence-with-migration (Yeaman and Whitlock 2011) showing that divergence patterns increasingly concentrate  during the late stages of divergence. Furthermore, in a later study Yeaman (2013) finds that the establishment bias  is not important when many loci underlie a selected trait, and this is true even during the early stages of divergence (for which Feder et al. (2012) found that the establishment bias is strongest). This is because  during the early stages of divergence, only a few loci manage to differentiate, and the probability that a mutation lands near any one of these few initially diverged loci is much smaller than the probability that it lands anywhere else in the genome. Based on this result, and disregarding stochastic loss, Yeaman (2013) concludes that clusters of differentiated loci cannot emerge in natural populations  during biologically realistic timescales unless other mechanisms are invoked that suppress recombination, such as genomic rearrangements. 

In summary, 
earlier studies are partly contradictory. Indeed, Yeaman and Whitlock (2011) demonstrate late formation of clusters of divergence while Feder et al.'s (2012) theoretical arguments predict the opposite. Moreover, and in contrast to both Yeaman and Whitlock (2011) and Feder et al. (2012), Yeaman's (2013) theoretical analysis excludes any possibility for the emergence of clusters of differentiation during divergence-with-migration, unless specific recombination-suppressor mechanisms are at work (factors that neither Yeaman and Whitlock (2011) nor Feder et al. (2012) included in their models). Critically, the existing theory relies on the assumption that any established local genomic divergence persists  indefinitely, or increases due to the accumulation of new beneficial mutations. But this is not the case in populations of finite size where genetic drift causes stochastic fluctuations of allele frequencies eventually leading to fixation of one (and the same) allele in diverging populations. Such a fixation event corresponds to loss of already established genomic divergence. 
This stochastic loss is of fundamental importance in all natural populations due to their finite sizes, and yet it is not known how loss influences patterns of genetic differentiation that arise during divergence-with-migration. Furthermore, because it is probably the case that in a majority of natural populations biological traits are controlled by a large number of loci (see, for example, reviews by Phillips (2008), and by Wagner and Zhang (2011)), then to interpret empirical data it is necessary to understand:  What are the genomic signatures of the process of stochastic loss during divergence-with-migration when many loci underlie the phenotype under selection? Does stochastic loss contribute to the formation of clusters of differentiated loci and, if so, how strongly? Finally, does the effect of this process change as divergence is ongoing?

To answer these questions we analyse a multi-locus model of divergence-with-migration, similar to that used by Yeaman and Whitlock (2011). By contrast to Yeaman and Whitlock (2011) and Yeaman (2013), we find that small, tightly-linked clusters of differentiated loci are necessary to initiate successful divergence under weak selection and high migration. Notably, these clusters form without invoking any specific mechanisms that reduce recombination. 
Furthermore, we show that clusters grow rapidly during the early stage of divergence, but shrink in size during the late stage. Under strong selection, by contrast, we find that clusters are not essential for divergence and that they instead form in the late stage of the process. Recall that increasing concentration in the late stage of the process has been reported by Yeaman and Whitlock (2011), but the formation and dynamics of clusters preceding this  stage that we find under weak selection and strong migration has, to our knowledge, not been reported or explained elsewhere. To explain these results  we analyse a two-locus model. We show that  the balance between `stochastic loss' and `gain' of local genomic divergence in finite populations is a  universal mechanism that governs the formation and temporal dynamics of clusters of differentiated loci. We stress that this a {\it universal}  mechanism because it is at work in all natural populations and unlike, for example, chromosomal rearrangements, it is not restricted to parts of the genome where specific recombination suppressors are active.

\newpage

\section{Materials and Methods}

\subsection{Multi-locus model}

We simulate a multi-locus model of divergence between two diploid  populations. The model is similar to that used in Yeaman and Whitlock (2011) (see also Supplementary file {\bf S1}). The two populations are assumed to occupy a pair of demes that are exposed to different environmental conditions, so that the phenotype is subject to opposing selection pressures in the two demes. We assume that in each deme (denoted by $k=1,2$) there is an optimal value $\theta^{(k)}$ for the phenotype. For simplicity, the two optima are assumed to be constant during time and symmetric around zero, so that $\theta^{(1)}=-\theta^{(2)}$.   

We assume that  the phenotype of an individual is determined by its diploid genotype at $L$ loci arranged on a single chromosome (but we also analyse the model with loci spread across two chromosomes, see {\bf Supplementary  information}).  In the model each allele is attributed an allele-effect size by which it contributes additively to the phenotype of an individual. In other words, the phenotype $z_i$ of individual $i$ equals the sum of allele-effect sizes  at the $L$ loci.  
We assume that the fitness $w^{(k)}_{i}$ of individual $i$ in population $k$ ($k=1,2$) depends on the phenotype $z_i$ of this individual as

\begin{equation}\label{eq:fitness}
w^{(k)}_{i}={\rm e}^{-\frac{(z_i-\theta^{(k)})^2}{2\sigma^2}}\,\,.
\end{equation}

\noindent Here  $\sigma$ is a parameter that determines the width of the distribution of the surviving phenotypes (Sadedin et al. 2009). When $\sigma$ is large, selection is weak, and vice versa. The selection parameter $\sigma$ is assumed to be constant during time and equal in the two populations. The fitness of an individual determines the contribution of this individual to the pool of offspring through soft fecundity selection. The  soft-selection assumption assures that the number of juveniles $N$ surviving to maturity in a given deme is constant over time, and we assume that it is equal in the two populations.   Generations are assumed to be discrete and non-overlapping.    
The lifecycle of individuals is modelled as follows. Virgin adults migrate to the opposite deme at a per-generation per-individual rate $m$. Migration is followed by random mating  locally within each population, recombination, and selection. Recombination is assumed to occur at a rate $r$ between adjacent loci, per-gamete, per-generation. Finally, mutations accumulate at a rate $\mu$ per allele, individual, generation. Each mutation is given a mutation-effect size by which it additively contributes to the effect size of the allele it lands on. Mutation-effect sizes are drawn randomly from a Gaussian distribution with a standard deviation $\sigma_\mu$, and a mean zero. To check whether the results are robust against the model for mutation-effect sizes, we also perform simulations in which mutation-effect sizes are drawn from an exponential distribution mirrored around zero, so that the mean mutation-effect size is zero. In these simulations the parameter of the exponential distribution is set to $\sqrt{2}/\sigma_\mu$ so that the variance of mutation-effect sizes is equal to $\sigma_\mu^2$. Finally, we note that the analysis in Martin and Lenormand (2006) of empirical data on fitness effects of mutations in different environments (data taken from various species) suggests that predictions of a model with a Gaussian fitness function and Gaussian distributed mutation-effect sizes  is in good agreement with a majority of the empirical data tested. 

\subsection{Parameter choices}
At the start of a simulation, all individuals at all loci are assumed to  have alleles of effect size zero.  
We set arbitrarily $\theta^{(1)}=-\theta^{(2)}=2$ (Table~I). In the majority of simulations, the number of loci $L$ is set to $L=100$, but we also test the model with $L=2000$ ({\bf Supplementary information}). The parameter $\sigma$ is chosen to account for weaker ($\sigma=4$) or stronger selection ($\sigma=2.5$). For further details on selection parameters, see Supplementary file {\bf S1}. To assess how the patterns are influenced by the local population size $N$, we contrast results obtained with  $N=1000$ and $N=200$. The migration rate $m$ is set to a  high value ($m=0.1$) that allows us to capture the signatures of migration under the chosen values of other model parameters. The recombination rate $r$ between a pair of adjacent loci is set to $r=0.0005$, or $r=0.001$ so that the first and the last locus in the genomic  region simulated (with $L=100$ loci) are at a recombination distance of about $0.05$, or $0.1$, respectively (but the distance is larger for $L=2000$). Note that $r=0.0005$ corresponds to about $5\cdot10^4$ base pairs assuming that recombination rate between two nearby base pairs is $10^{-8}$. The mutation rate $\mu$ per generation, allele, locus, individual is chosen so that mutations that influence an individual's phenotype occur infrequently ($\mu=2\cdot10^{-5}$). Finally, the variance of mutation-effect sizes $\sigma_\mu^2$  is set to a small value ($\sigma_\mu=0.05$) so that the square root of the total variance over all adaptive loci ($\sqrt{L} \sigma_\mu$) is smaller than the distance between the optimal trait values $\theta^{(1)}-\theta^{(2)}$. For the parameters set here and assuming that the whole genome region simulated acts as a single locus (total recombination rate is equal to zero), this means that it requires, on average, about $4$ adaptive steps for the populations to reach their optimal traits (taking into account diploidy). Otherwise, if the distance between the optima is equal to, or lower than $\sqrt{L} \sigma_\mu$, perfect adaptation in both populations can, by chance, be attained in a single adaptive step, which we consider to be an unlikely scenario in natural populations. 

The model is simulated for a large number of generations (up to $10^5$) to allow the populations to come close to their local optima and stabilise. At intervals of fifty generations, we measure the extents of local and total genomic divergence as follows. The extent of local genomic divergence $D_{l}$ at  locus $l$ is estimated  as twice the difference between the allele-effect size of the most frequent alleles at this locus in the two populations. The factor $2$ is used because the population is diploid. Our measure of local genomic divergence divided by $2$ corresponds to the measure $d$ used by Yeaman and Whitlock (2011).  We approximate the extent of total genomic  divergence $D$ in a given generation by summing the extents of local genomic divergence at all loci in this generation.  For our parameters, perfect adaptation in both populations corresponds to the total genomic divergence equal to the difference between the local optima $\theta^{(1)}-\theta^{(2)}=4$. As an alternative to the measure $D_l$ of the extent of local genomic divergence at locus $l$, one can use twice the average allele-effect size at this locus (and the sum over all loci would correspond to the average total extent of divergence). We note that the two measures of divergence (one based on the most frequent alleles, and the other on the average allele-effect sizes in the two populations)  give rise to qualitatively the same patterns of divergence (see below).  However, because the measure $D$ is directly comparable to the measure $d$ used by Yeaman and Whitlock (2011), we present most of our results in terms of this measure. Divergence patterns presented similarly to Yeaman and Whitlock (see their Figure 2), allow one  to inspect visually each individual realisation of the model and to evaluate roughly whether clusters of divergence are formed and, if yes, what is their typical size. Here, however, we complement such a visual inspection by measuring correlations of local extents of genomic divergence at pairs of loci as a function of their recombination distance. This allows us to capture the extent of similarity of differentiation at pairs of loci at various recombination distances in a given generation. Note that when a cluster forms, the extents of divergence at loci within the cluster are expected to be more correlated than the extents at loci outside of the cluster. Therefore, if a cluster is formed, its size (that is, the recombination distance it spans) is expected to be captured by the recombination distance at which the correlation function decays to values close to zero. For each parameter set we run $90$ independent simulations (unless otherwise noted) to evaluate the effect of stochastic fluctuations on the extents of local and total divergence.

\subsection{Two-locus model}
To understand the mechanisms at work in the multi-locus model presented above, we analyse a two-locus model. In particular, we use two versions of a two-locus model. One is an {\it establishment model} (similar to the models used by Feder et al. 2012, and by Yeaman 2013), and the other is a (novel) {\it gain-loss model}. These two are briefly explained next (but see also Supplementary files {\bf S2}-{\bf S3}).
 
As noted in the introduction, earlier theory of divergence-with-migration focuses on evaluating the importance of the establishment bias for  the evolution of genetic architectures during divergence-with-migration (Yeaman and Whitlock 2011; Feder et al. 2012; Yeaman 2013). Recall that the establishment bias here means that the probability of establishment of a new mutation is larger for the mutation landing closer to an already diverged locus than further away. While the establishment bias can be significant when very few loci underlie a selected trait (Yeaman and Whitlock 2011; Feder et al. 2012), this is not true when many loci underlie the trait (Yeaman 2013). To check whether this finding of Yeaman (2013) holds true for the parameter values used in this study (see {\bf Parameter choices}), we employ the establishment model  analysing  whether or not there is a range of recombination distances around an already diverged locus  such that a successful establishment of a new mutation is more likely within this range than outside of it (see below). In this study {\it a successful establishment} of a mutation means that the mutant allele is most common (frequency $>50\%$)  in the deme where it is advantageous (cf. Yeaman and Otto 2011). In the establishment model, one locus is differentiated at the outset, the other is not (Supplementary file {\bf S2}). We analyse the establishment probability of new mutations at the undifferentiated locus, varying its recombination distance from the differentiated one. 
Using these results, we compare the probability that a new mutation lands and establishes within a genomic region of a given size around the diverged locus, to that outside of this region (as suggested earlier by Yeaman (2013)). When the mutation rate per locus is equal for all loci (as we assume here  and in accordance with Yeaman 2013), the ratio between the two probabilities is independent of the mutation rate, and it is equal to the integral of the establishment probability over recombination distances within a region relative to the integral of the establishment probability over recombination distances outside of this region (but within the total genomic region considered). If this ratio is greater than unity, we can immediately deduce that a cluster of divergence is likely to be formed. Otherwise, a new mutation may establish at any recombination distance from the differentiated locus. By contrast to Yeaman (2013), we argue, however, that in this case we cannot draw a final conclusion about cluster formation because, once established, any local genomic divergence is subject to two  competing processes. One process is stochastic loss that occurs due to random genetic drift  in populations of finite size, resulting in fixation of a single allele at a given locus in both populations. The other process is  the gain of additional local genomic divergence that occurs due to the influx of new mutations followed by their successful establishment. The ratio of the rates at which these two local processes operate (hereafter referred to as {\it balance}) determines whether or not divergence established in a genomic region at a given divergence stage will make a lasting contribution to overall differentiation. The balance between these two processes at a given locus depends on the distance of this locus from other diverged loci in the genome, as well as the strength of local and total genomic divergence. To understand this dependence, we use the two-locus gain-loss model.

In this model,  both loci are assumed to have established divergence. One locus is assumed to be weakly differentiated with the extent of divergence $D_{\rm w} $ corresponding to the establishment of one mutation beneficial in the first population, and one mutation beneficial in the second one. We set  the allele-effect sizes at this locus  to $Y_{\rm w}=\sigma_\mu$, and $-Y_{\rm w}$, so that $D_{\rm w}=4\sigma_\mu$ (for further details on this choice, see Supplementary file {\bf S3}). 
The second locus is assumed to have stronger divergence $D_{\rm s}$ ($> D_{\rm w})$ with allele-effect sizes $Y_{\rm s}$ ($>Y_{\rm w}$), and $-Y_{\rm s}$ (and so $D_{\rm s} =4Y_{\rm s}$). We vary the value of $Y_{\rm s}$ in the simulations to mimic different stages of divergence (see Supplementary file {\bf S4} for details on the initial conditions in these simulations). Using this model, we estimate the rate of local gain (product of the rate at which a mutation lands at a locus and the rate at which this mutation establishes successfully conditional on it landing at the locus), and the rate of local loss at the two differentiated  loci. When the rate of loss  at a locus locus  is larger than the rate of gain at this locus in a given stage of divergence,  this locus is  unlikely to make a sustained contribution to overall divergence.  Otherwise, the opposite is true. In what follows we explain our method for estimating the rates of gain and loss using this model.

In the limit of rare mutations ($2\mu N\ll 1$) and when the two loci are at a recombination distance $r_j=j r$ ($j=1,\ldots,L$), the rate of local gain $\lambda_{\rm G,w}(r_j)$ at the weakly diverged locus  is equal to the product of the probability that a locally beneficial mutation lands at this locus ($2\mu N$), and the probability  $p_{\rm G,w}(r_j)$ that it establishes at this locus (conditional on the mutation landing at the locus)

\begin{equation}\label{eq:gain}
\lambda_{\rm G,w}(r_j)=2\mu Np_{\rm G,w}(r_j)\,\,.
\end{equation}

\noindent Substituting subscripts $w$ in Eq.~(\ref{eq:gain}) by $s$ we obtain the corresponding expression for the rate of local gain at the more strongly diverged locus.
In the limit of $\lambda_{\rm G,w}(r_j)\ll 1$, the time to local gain at the weakly diverged locus is approximately exponentially distributed with mean $\lambda_{\rm G,w}^{-1}(r_j)$ (and similarly for the more strongly diverged locus).

We estimate the rates of gain at the two loci using two separate sets of simulations. In one set we assume that a mutation of a fixed mutation-effect size  $\epsilon=\sigma_\mu$ (as in the establishment model, see Supplementary file {\bf S2})  lands at the weakly diverged locus immediately after the initialisation of the system (the initial condition is explained in detail in Supplementary file {\bf S4}). In the other set of simulations we assume that the mutation lands on the more strongly diverged locus (all other settings are the same as in the former set of simulations). Further mutations are thereafter neglected. We further assume that in the former case the mutation lands in the first population (where it is beneficial) on an allele of effect size $Y_{\rm w}$ (and similarly on $Y_{\rm s}$ in the latter case). For these settings, the mutant allele is advantageous over both alleles at the locus prior to the mutation, and hence it will promote the local extent of divergence upon a successful establishment. We use a similar method to that explained in the establishment model (see above) to  estimate the establishment probabilities $p_{\rm G,w}(r_j)$, and $p_{\rm G,s}(r_j)$ at the weakly and at the more strongly diverged locus, respectively. Finally, we use Eq.~(\ref{eq:gain}) to estimate the rates of gain at the two loci.

In addition to the rates of gain, we estimate the rates of local loss at the two loci starting from the same initial condition as in the simulations described above, but now neglecting mutations. Each simulation is run under drift, selection, migration and recombination until one or the other locus experiences loss  of divergence or until a predetermined maximum time ($T_{\rm m}$) expires. As explained above, {\it loss of divergence} means that a locus becomes monomorphic due to fixation of one allele in both populations. Using simulations, we first estimate the probabilities $p_{\rm L,s} (r_j|T_{\rm m})$ and $p_{\rm L,w} (r_j|T_{\rm m})$ that the first loss event occurs at the more strongly or at the  weakly diverged locus, respectively, conditional on it occurring before $T_{\rm m}$. Here $r_j$ denotes the recombination distance between the two loci (see above). Second, we estimate the average time $\langle t_{\rm L} (r_j |T_{\rm m})\rangle$ until the first loss event based on simulations in which a loss has occurred by the maximum time $T_{\rm m}$.  Using these data, we estimate the rates of loss $\lambda_{\rm L,s} (r_j)$  and $\lambda_{\rm L,w} (r_j)$ at the more strongly and at the weakly diverged locus, respectively, based on the following considerations. In the limit of $\lambda_{\rm L,s} (r_j)\ll 1$, the time to loss at the more strongly differentiated locus is approximately exponentially distributed with  mean $\lambda_{\rm L,s} ^{-1}(r_j)$ (and similarly for the weakly diverged locus). Furthermore, in the limit of  $\lambda_{\rm L,s} (r_j)\ll 1$, $\lambda_{\rm L,w}(r_j)\ll 1$, the time until the first loss event (either at the first or at the second locus) is approximately exponentially distributed with mean $(\lambda_{\rm L,s} (r_j)+\lambda_{\rm L,w} (r_j))^{-1}$. Therefore the probability  $p_{\rm L}(r_j|T_{\rm m})=p_{\rm L,s} (r_j|T_{\rm m})+p_{\rm L,w} (r_j|T_{\rm m})$ that the first loss event occurs  either at the more strongly or at the weakly diverged locus by the time $T_{\rm m}$ is given by

\begin{equation}\label{eq:loss}
p_{\rm L}(r_j|T_{\rm m})=1-{\rm e}^{-\Bigl[\lambda_{\rm L,s} (r_j)+\lambda_{\rm L,w} (r_j)\Bigr]T_{\rm m}}\,\,.\end{equation}

\noindent Finally, we find that the average time $\langle t_{\rm L}(r_j |T_{\rm m})\rangle$ to the first loss event, conditional on the loss occurring by the time $T_{\rm m}$, can be expressed in terms of $\lambda_{\rm L,s} (r_j)$, $\lambda_{\rm L,w}(r_j)$, $p_{\rm L}(r_j|T_{\rm m})$, and $T_{\rm m}$ as follows:

\begin{equation}
\langle t_{\rm L}(r_j |T_{\rm m})\rangle=\frac{1}{\lambda_{\rm L,s} (r_j)+\lambda_{\rm L,w} (r_j)}-T_{\rm m}\frac{1-p_{\rm L}(r_j|T_{\rm m})}{p_{\rm L}(r_j|T_{\rm m})}\,\,.
\end{equation}

\noindent Since 

\begin{equation}\label{eq:ratio1}
\frac{p_{\rm L,s} (r_j|T_{\rm m})}{p_{\rm L,w} (r_j|T_{\rm m})}=\frac{\lambda_{\rm L,s} (r_j)}{\lambda_{\rm L,w} (r_j)}\,,\end{equation} 

\noindent we obtain

\begin{align}
\lambda_{\rm L,w} (r_j)=\frac{p_{\rm L,w} (r_j|T_{\rm m})}{p_{\rm L}(r_j|T_{\rm m})}\Bigl(\langle t_{\rm L}(r_j |T_{\rm m})\rangle+T_{\rm m}\frac{1-p_{\rm L}(r_j|T_{\rm m})}{p_{\rm L}(r_j|T_{\rm m})}\Bigr)^{-1}\,\,.\label{eq:lambda_s}
\end{align}

\noindent We use Eq.~(\ref{eq:lambda_s}) to estimate the rate of loss  $\lambda_{\rm L,w} (r_j)$ given the probabilities  $p_{\rm L,s} (r_j|T_{\rm m})$ and $p_{\rm L,w} (r_j|T_{\rm m})$ and the average time $\langle t_{\rm L}(r_j |T_{\rm m})\rangle$ that we obtain using simulations. The rate of local loss at the more strongly diverged locus is obtained by combining Eqs.~(\ref{eq:ratio1})-(\ref{eq:lambda_s}).

Note that $\langle t_{\rm L} (r_j |T_{\rm m})\rangle$ is not defined if no loss occurs by the maximum time $T_{\rm m}$ set in the simulations. 
To avoid such cases, $T_{\rm  m}$ has to be long enough ($T_{\rm m}\gg(\lambda_{\rm L,s} (r_j)+\lambda_{\rm L,w} (r_j))^{-1}$) to assure that loss occurs by this time with a high enough probability. Since we do not know the rates $\lambda_{\rm L,s} (r_j)$, $\lambda_{\rm L,w} (r_j)$ in advance, $T_{\rm m}$ has to be chosen. Here we set it to a large value $T_{\rm m}=10^5$, 
because this allows us to compare the timescales of local loss and gain for other parameter values used in this study. Indeed,  when the mutation rate is $\mu=2\cdot 10^{-5}$ (as in Fig.~1), this value of $T_{\rm m}$ corresponds to twice the average waiting time until a mutation establishes successfully at a neutral locus in an isolated population. Hence $T_{\rm m}$ is larger than the average time to local gain at any locus under selection (inverse of the rate of gain, see above). Importantly, in situations when no loss occurs by this time, we immediately deduce that the rate of local loss is much smaller than the sum of rates of gain at the weakly and at the more strongly diverged locus (according to Eq.~(\ref{eq:loss})).

\newpage
\section{Results}
Under weak selection (Fig.~1{\bf A}, {\bf B}) we find an initial phase of roughly homogeneous divergence over the adaptive loci with the pairwise correlation of divergence being independent of the recombination distance between loci (Fig.~1{\bf B}). In this phase, divergence at any one locus is highly transient and the total extent of divergence is very low. After a waiting time of about ten thousand generations (in this particular realisation, but see other examples in Fig.~S1), groups of closely linked loci establish divergence. This initiates rapid formation of a cluster of divergence that extends in size and immediately promotes the advance of phenotypic adaptation.   At about half way towards perfect adaptation ($D\approx 2$), the cluster of diverged loci attains a maximum size (Figs.~1{\bf A},~S1), with an average of around $15$ loci (Fig.~1{\bf B}).  Thereafter, the cluster shrinks in size, but most of the cluster still remains after $10^5$ generations. 

Under strong selection, the initial phase is also characterised by roughly uniformly distributed divergence (Fig.~1{\bf C}, {\bf D}). The build-up of divergence is, as expected, much faster under strong selection, and the formation of a cluster is not necessary to initiate population divergence. Even so, when approximately perfect adaptation is attained under strong selection ($D\ge 4$), divergence starts to concentrate, resulting in formation of a cluster of divergence. Note that, comparing to weak  selection, a cluster under strong selection starts forming in a much later stage of divergence in terms of the value of $D$, but sooner in terms of the number of generations after the start of divergence ($2000$ rather than $10 000$ generations in the particular realisations shown in Fig.~1{\bf C}, and {\bf A}, respectively). Note also that the divergence patterns obtained using exponentially distributed mutation-effect sizes, with otherwise the same settings as those in Fig. 1, do not qualitatively differ from the patterns shown in Fig. 1 (see Fig.~S9). The same is true if local extents of genomic divergence are measured using the average allele-effect size instead of the measure used in Fig.~1 (see Figure S10).

To investigate further the differences in divergence patterns obtained for weak and strong selection (Fig.~1{\bf A} and {\bf C}, respectively), we estimate the establishment probabilities of a new mutation as a function of the recombination distance from an already diverged locus (Supplementary Figure~S2). Next, we compare the probability that a mutation lands and establishes outside of a genomic region around the diverged locus to the probability that it lands and establishes inside of this region (Supplementary Figure~S3). 
For a small genomic region surrounding the diverged locus we find that the probability of landing and establishment outside of the region is much larger than the corresponding probability inside of the region. Furthermore, even when the region accounts for $50\%$ of the whole genomic region simulated ($100$ loci), the corresponding probability outside is only slightly less than the probability inside of the region.
 These findings (consistent with Yeaman 2013) are  true both for weak and strong selection, suggesting that  the establishment bias is too weak to cause the formation of clusters, especially the tightly-linked ones observed under weak selection in our multi-locus simulations.   
Consequently, we need an additional mechanism to explain the emergence of a cluster under weak selection.

Our gain-loss model helps to understand the progress of cluster formation. In this two-locus model both loci already have some divergence established, but one has diverged more strongly ($D_{\rm s}\ge 0.4$) than the other ($D_{\rm w}=0.2$). 
First, analysing the rate of gain at the two loci in an initial stage of the divergence process  ($D=0.6$, $15\%$ of the value corresponding to perfect divergence), we find that the rate is marginally larger at the more strongly diverged locus (Fig.~2{\bf A}, {\bf B}). For both loci, the rate of gain of new genetic differentiation is higher when the two loci are at a smaller recombination distance but this distance dependence  is weak for either weak or strong selection (Fig.~2{\bf A}, {\bf B}). Recall that the rate of gain is the product of the mutation rate and the establishment  probability of  a new mutation conditional on it landing locally in the genome. Therefore, the rate of gain depends weakly on the recombination distance between the loci due to a weak bias in the establishment probability discussed above.

Second, we analyse the risk of loss of divergence by stochastic processes that may eliminate variation in either of the two loci. Under weak selection, we show that the rate of loss of divergence at the more diverged locus is small and depends only weakly on the recombination distance between the two loci (Fig.~2{\bf C}, squares).  However, at the less diverged locus the rate of local loss increases rapidly with recombination distance from the other locus (Fig.~2{\bf C}, circles). Comparing the rate of gain and loss at the weakly diverged locus, we find a critical recombination distance from the more strongly diverged locus above which loss is on average faster than gain. Divergence established at a less diverged locus above this critical distance is unlikely to make a lasting contribution to the overall divergence. In fact, in this initial stage of divergence, the rate of loss is marginally larger than the rate of gain already at one locus distance from the more differentiated locus (where distance is scaled by the recombination rate between a pair of adjacent loci). As a consequence, in most cases a weakly diverged locus adjacent to a more strongly diverged one fails to contribute to further divergence.  However, occasionally, as a matter of chance and after a shorter or longer waiting time, divergence is gained at the weakly diverged locus. 
By contrast, under strong selection no loss of divergence occurred during  $200$ simulations at either of the two loci,  and so in this case the rate of local loss is much smaller than the rate of local gain for all between-locus recombination distances that we analysed (Fig.~2{\bf D}).

After the initial stage, the divergence rapidly continues to increase both under weak and strong selection (Fig.~1). This alters the balance between rates of loss and gain. From the two-locus model we find that under weak selection, the critical recombination distance between the two loci below which there is a net gain of new divergence, initially increases leading to an overall increase in divergence and cluster size (compare positions of arrows in Fig.~3{\bf A}-{\bf D}). Half way towards perfect local adaptation ($D_{\rm s}\approx 2$) the critical distance starts to decrease. Consequently, under weak selection and as $D_{\rm s}$ increases up to $D_{\rm s}\approx 2$, the cluster grows reaching the size of about twenty loci (two times the maximum critical scaled distance of $10$ loci in Fig. 3{\bf B}). This largely corroborates the finding of the multi-locus model where the maximum cluster size attained during divergence is about $15$ loci on average. Thereafter, the two-locus model predicts that the cluster will shrink in size and this is also observed in the multi-locus model  (Fig.~1{\bf A}). Under strong selection, by contrast, there is a continued net gain of divergence until populations approach perfect adaptation ($D\approx 4$, Fig.~4{\bf A}). At this stage, both populations have at their disposal gene variants that combined together give rise to locally perfectly adapted individuals. However, stochastic loss of divergence becomes increasingly important for alleles of small effect that are loosely linked to loci with stronger divergence (Fig.~4{\bf B}). Adaptation is maintained by gain at closely-linked loci, leading to increasing clustering.

The key features of cluster formation early in divergence and decrease in cluster size later in divergence are retained for higher recombination rate, intermediate selection, lower population size and larger genomic region (Supplementary Figures~S4-S8).
\newpage
\section{Discussion}

Reproductive isolation between populations is most efficient when many small barriers to gene flow are formed throughout the genome  (Barton 1983; Coyne and Orr 2004). 
Otherwise, linkage to barrier loci may be insufficient to prevent gene flow over a large part of the genome (Barton and Bengtsson 1986). Thus, the genomic distribution and effect sizes of loci underlying local adaptation are critical to understanding the origin of reproductive isolation in models of ecological speciation with gene flow (Nosil 2012; Seehausen et al. 2014).

Models of divergence are attractive in the sense that they suggest different mechanisms by which the barrier effects of single adaptively divergent loci may be enhanced so that the total barrier increases and the genomic region affected broadens (Via 2009; Feder et al. 2012). However, recent simulation studies suggest that local barrier effects are enhanced only extremely late in the process  (Yeaman and Whitlock 2011), or that they are unlikely to be of any biological relevance unless interacting with chromosomal inversions and other genomically-localised mechanisms that reduce recombination (Feder et al. 2013; Yeaman 2013). 
In contrast to these conclusions, we here show that a specific mechanism that suppresses recombination is not necessary for clusters of differentiation to form. Moreover, we show that under weak selection and strong migration, the emergence of a concentrated genetic architecture is indispensable for phenotypic divergence to evolve. These findings are, to our knowledge, new and contribute to explaining observed empirical patterns, as discussed below.

The reason we  detect clusters of differentiation despite the fact that the  establishment bias (referred to as `divergence hitchhiking' by some authors (Feder et al. 2012; Yeaman 2013)) is too weak to support clustering is because a mechanism beyond the establishment bias is at work. In contrast to the establishment probability, the rate of loss of differentiation at a weakly diverged locus depends strongly on the recombination distance to a locus of stronger effect, and so the balance between loss and gain of small extents of local differentiation also depends strongly on the recombination distance. This balance between loss and gain is the key mechanism underlying the formation of clusters of divergence. 

Our results show that when new locally beneficial mutations are under weak selection and migration between the diverging populations is frequent, tightly-linked clusters of differentiated loci are a prerequisite for initialisation of successful phenotypic divergence.
 The initialisation occurs after a waiting time that is, on average, longer for parameter settings giving rise to smaller clusters, i.\,e.  weaker selection, smaller variance of mutation-effect sizes, smaller population sizes, higher migration rate.   When selection for locally beneficial mutations is sufficiently strong, however, we find rapid phenotypic divergence that precedes cluster formation.

Our two-locus analysis of the interplay between loss and gain of local genomic divergence is highly consistent with the results of the multi-locus modelling. 
A key idea is that any locus that has established divergence may either risk a stochastic loss of divergence (similar to the idea of `transient divergence' (Yeaman 2015)) or gain from additional beneficial mutations. Consequently, a diverged locus is, at a given stage, unlikely to make a lasting contribution to the overall divergence if the rate of local loss is larger than the rate of local gain. There is a critical recombination distance from the focal locus above which local loss is faster than local gain. This distance depends on the selection strength and it varies over the time-frame of the divergence process. Under weak selection, the critical distance increases early in the divergence process but, about half-way to perfect adaptation, it starts to decrease. The latter effect arises because, after about half-way to perfect adaptation for the parameter values we tested, a weakly differentiated locus at a given recombination distance from other differentiated loci contributes proportionately by a very small amount to the overall extent of divergence and to the reduction of gene flow between the populations. This contribution becomes smaller as the total extent of divergence increases beyond a point corresponding to about half-way to perfect adaptation. Consequently, as divergence progresses above this point, the rate of gain of new differentiation at a given weakly differentiated locus decreases, and the rate of loss increases. Therefore, the ratio between the rate of local loss and the rate of local gain increases, resulting in shrinking of a cluster over time. For a similar reason genetic architectures concentrate in late stages of the divergence process also under stronger selection (or weaker migration) but this occurs later in the divergence process. In particular, under strong selection considered here (see also Yeaman and Whitlock 2011), diverging populations attain almost perfect adaptation before clustering of the genetic architecture starts. The dynamics of clusters obtained under our multi-locus simulations is, therefore, consistent with the main predictions of the two-locus gain-loss model in different stages of divergence.  Notably, because our analysis contrasts the effects of loss and gain locally in the genome, the consequences of the balance between these two effects for the size of a cluster (that is, the recombination distance it spans) in multi-locus models of divergence is independent of the number of selected loci, provided that this number is large.    

We find that the cluster size emerging in our model is well characterised by the correlation function describing the similarity in extents of divergence in pairs of loci in relation to their recombination distance. When clusters are formed, the correlation decreases with increasing recombination distance between the loci, reaching approximately zero at the cluster margin. This measure is closely related to measures of linkage disequilibrium (McVean 2002; Eriksson and Mehlig 2004; Schaper et al. 2012) that are frequently used in empirical studies (Smadja and Butlin 2011; Martin et al. 2013). 

Apart from the findings discussed above, we also find that when multiple tightly-linked clusters emerge during divergence (see an example in Fig. S1{\bf A}), the clusters compete with each other for gaining new differentiation (or against losing the differentiation they have established). The dynamics of such a competition can be investigated by a gain-loss model similar to that analysed here, but with more than two loci included and focusing on gain and loss of differentiation at individual clusters, each of which contains multiple loci.

Some modelling studies have earlier considered the loss of divergence. Using a single-locus model, Yeaman and Otto (2011) found that less diverged loci have a smaller persistence time than more diverged ones. A single-locus analysis is, however, insufficient to explain clustering, because it is not only the extent of local divergence that matters but also the extent of divergence at other diverged loci and their linkage. In addition, it is not the persistence time {\it per se} that matters but, as shown here, a balance between loss and gain processes which operates differently in different stages of divergence. In a recent study, Aeschbacher and B\"urger (2014) analysed a two-locus continent-island model of divergence, deriving an approximation for the mean extinction time of a mutation at some recombination distance from a diverged locus. Comparing the mean extinction time at a linked locus with that at an unlinked one, they showed that the mean extinction time is shorter when linkage is looser. However, this comparison may not be relevant for the patterns of divergence because, as we show, the balance between local loss and gain shifts over the timescale of the process. 

Due to our upper limit of $10^5$ generations, we do not capture the final fate of the clusters. Yeaman and Whitlock (2011) suggested that a pair of populations undergoing divergence-with-migration will eventually differ at a single locus, and our results seem to corroborate this conclusion. We note, however, that  other factors, such as the evolution of habitat choice or assortative mating, may reinforce isolation (Thibert-Plante and Gavrilets 2013). These processes are likely to prevent clustering in late stages of divergence by reducing gene flow between populations and introducing additional mechanisms at work  (Cruickshank and Hahn 2014).

Empirical studies report either little evidence of genomic clustering (Soria-Carrasco et al. 2014), or strong evidence for generally small clusters  (Jones et al. 2012), or for two orders of magnitude larger clusters (Ellegren et al. 2012). This large variation in cluster size may hint that different mechanisms are involved  (Seehausen et al. 2014), including those that reduce recombination (Yeaman 2013). Theory suggests that inversions might be more important than other recombination suppressors because they work specifically in heterozygotes, rather than generally suppressing recombination (Otto and Lenormand 2002). However, there is evidence for fine-scale variation in recombination rates that is also likely to contribute to heterogeneous patterns (Burri et al. 2015). 

In general, the extent of migration between the diverging populations is an important factor shaping the genetic architectures evolving during divergence (Feder et al. 2013, Seehausen et al. 2014). For example, in a recent study by Marques et al. (2016), it has been shown that genetic differentiation between sympatric races of threespine sticklebacks is concentrated in the genome, occurring over few very short genomic regions on only two chromosomes (see their Fig. 3C). By contrast, many more differentiated genes and chromosomes are detected between essentially allopatric races of this species, suggesting a roughly uniformly distributed differentiation (see Fig. 3D in Marques et al. 2016). These results seem consistent with the predictions of our model comparing high and small migration rate between the diverging populations. However Marques et al. (2016) also suggest that the diverging populations they analysed probably had some amount of standing genetic variation at the time they were introduced to the sites examined. The role of standing genetic variation, however, is not examined by our current model, and we find this to be an important future avenue.

In summary, of the different mechanisms potentially contributing to the formation of local barriers to gene flow, the balance between the processes of local loss and local gain that we proposed here is, to our knowledge, the only universal mechanism that promotes concentrated genetic architecture under strong gene flow, without suppressing recombination. We show that the number of loci in a cluster is smaller under weaker selection, smaller mutation-effect sizes, smaller population size, and stronger recombination. All of these parameters are likely to vary among species, and among populations within species. Furthermore, our model predicts systematic changes in cluster size during divergence. Thus, the balance between loss and gain of local genomic divergence potentially explains much of the observed variation in genomic architectures emerging during divergence-with-migration and leads to testable predictions about the causes of this variation.

\newpage

\begin{acknowledgments}
This work was supported by a Linnaeus-grant from Vetenskapsr{\aa}det and Formas (http://www.cemeb.science.gu.se),  as well as by additional grants from Vetenskapsr\aa det, 
and from the G\"oran Gustafsson Foundation for Research in Natural Sciences and Medicine. RKB also thanks the support from Tage Erlander and Waernska Guest Professorships, as well as the Natural Environment Research Council.  We thank Anja Westram for very helpful comments on an early draft of this manuscript. 
F. Wang and P. Tr\"askelin contributed to the analysis of a model for the evolution of haplotypes under selection in a single well-mixed population.
\end{acknowledgments}

\section*{Author contributions} 
MR, KJ, RKB and BM designed the study. MR and AE executed and analysed the two-locus establishment model. MR executed and analysed the multi-locus model, as well as the two-locus gain-loss model. MR, KJ, RKB and BM drafted the manuscript. All authors contributed to revisions.

\section*{Competing financial interests}
Authors declare no competing financial interests.
\section{References}

\begin{hangparas}{.25in}{1}
Aeschbacher, S., and R. B\"urger. 2014.  The effect of linkage on establishment and survival of locally beneficial mutations. Genetics 197(1): 317--336.
\end{hangparas}

\begin{hangparas}{.25in}{1}
 Barton, N. H.  1983. Multilocus clines. Evolution 37(3): 454--471.
\end{hangparas}

\begin{hangparas}{.25in}{1}
Barton, N. H., and B. O. Bengtsson. 1986. The barrier to genetic exchange between hybridising populations. Heredity 57(3): 357--376.\end{hangparas}

\begin{hangparas}{.25in}{1}
Bengtsson, B. O.  1985. The flow of genes through a genetic barrier. Pp 31--42 {\it in} Greenwood, J. J., P. H. Harvey,  and M. Slatkin, eds. Evolutionary essays in honour of John Maynard Smith. Cambridge Univ. Press, Cambridge.
\end{hangparas}

\begin{hangparas}{.25in}{1}
 Burri, R. et al. 2015. Linked selection and recombination rate variation drive the evolution of the genomic landscape of differentiation across the speciation continuum of {\em Ficedula} flycatchers. Genome Res. 25(11): 1656--1665.
\end{hangparas}

\begin{hangparas}{.25in}{1}
Coyne, J. A., and H. A. Orr. 2004. Speciation. Sinauer Associates, Sunderland, MA.
\end{hangparas}

\begin{hangparas}{.25in}{1}
 Cruickshank, T. E., and M. W. Hahn. 2014.  Reanalysis suggests that genomic islands of speciation are due to reduced diversity, not reduced gene flow. Mol. Ecol. 23(13): 3133--3157.
\end{hangparas}

\begin{hangparas}{.25in}{1}
Ellegren, H. et al. 2012. The genomic landscape of species divergence in {\it Ficedula} flycatchers. Nature  491(7426): 756--760.
\end{hangparas}

\begin{hangparas}{.25in}{1}
Eriksson, A., and B. Mehlig. 2004.  Gene-history correlation and population structure. Phys. Biol. 1(4): 220--228.
\end{hangparas}

\begin{hangparas}{.25in}{1}
Feder, J. L., S. M. Flaxman, S. P. Egan, A. A. Comeault, and P. Nosil. 2013.  Geographic
mode of speciation and genomic divergence. Annu. Rev. Ecol. Evol. Syst. 44: 73--97.
\end{hangparas}

\begin{hangparas}{.25in}{1}
Feder, J. L., R. Gejji, S. Yeaman, and P. Nosil. 2012. Establishment of new mutations under divergence and genome hitchhiking. Philos. Trans. R. Soc. Lond. B Biol. Sci. 367(1587): 461--474.
\end{hangparas}

\begin{hangparas}{.25in}{1}
Feder, J. L., and P. Nosil. 2010. The efficacy of divergence hitchhiking in generating genomic islands during ecological speciation. Evolution 64-6: 1729--1747.
\end{hangparas}

\begin{hangparas}{.25in}{1}
Flaxman, S. M., J. L. Feder, and P. Nosil. 2013. Genetic hitchhiking and the dynamic buildup of genomic divergence during speciation with gene flow. Evolution 67(9): 2577--2591.
\end{hangparas}

\begin{hangparas}{.25in}{1}
 Jones, F. C. et al. 2012. The genomic basis of adaptive evolution in threespine sticklebacks. Nature  484(7392): 55--61.
\end{hangparas}

\begin{hangparas}{.25in}{1}
 Martin, S. H. et al. 2013.  Genome-wide evidence for speciation with gene flow in {\it Heliconius} butterflies. Genome Res. 23(11): 1817--1828.
\end{hangparas}

\begin{hangparas}{.25in}{1}
 Martin, G. and Lenormand, T. 2006.  The fitness effect of mutations across environments: a survey in light of fitness landscape models. Evolution 60(12): 2413--2427.
\end{hangparas}

\begin{hangparas}{.25in}{1}
Marques, D. A. et al. 2016. Genomics of rapid incipient speciation in sympatric threespine stickleback. PLoS Genet. 12(2): e1005887.
\end{hangparas}

\begin{hangparas}{.25in}{1}
 McVean, G. A.  2002. A genealogical interpretation of linkage disequilibrium. Genetics 162(2): 987--991.
\end{hangparas}

\begin{hangparas}{.25in}{1}
Nosil, P. 2012. Ecological speciation. Oxford University Press, Oxford.
\end{hangparas}

\begin{hangparas}{.25in}{1}
 Otto, S. P. and T. Lenormand. 2002. Resolving the paradox of sex and recombination. Nat. Rev. Genet. 3(4): 252--261.
\end{hangparas}

\begin{hangparas}{.25in}{1}
Phillips, P. C. 2008. Epistasis- the essential role of gene interactions in the structure and evolution of genetic systems. Nat. Rev. Genet. 9: 855--867.
\end{hangparas}

\begin{hangparas}{.25in}{1}
 Sadedin, S., J. Hollander, M. Panova, K. Johannesson, and S. Gavrilets. 2009. Case studies and mathematical models of ecological speciation. 3: Ecotype formation in a Swedish snail. Mol. Ecol. 18(19): 4006--4023.
\end{hangparas}

\begin{hangparas}{.25in}{1}
Savolainen, O., M. Lascoux, and J. Meril\"a. 2013.  Ecological genomics of local adaptation. Nat. Rev. Genet. 14(11): 807--820.
\end{hangparas}

\begin{hangparas}{.25in}{1}
 Schaper, E., A. Eriksson, M. Rafajlovi\'c, S. Sagitov, and B. Mehlig. 2012. Linkage disequilibrium under recurrent bottlenecks. Genetics 190(1): 217--229. 
\end{hangparas}

\begin{hangparas}{.25in}{1}
Seehausen, O. et al. 2014. Genomics and the origin of species. Nat. Rev. Genet. 15(3): 176--192.
\end{hangparas}

\begin{hangparas}{.25in}{1}
Smadja, C. M. and R. K. Butlin. 2011.  A framework for comparing processes of speciation in the presence of gene flow. Mol. Ecol. 20(24): 5123--5140.
\end{hangparas}

\begin{hangparas}{.25in}{1}
Soria-Carrasco, V. et al. 2014. Stick insect genomes reveal natural selection's role in parallel speciation. Science 344(6185): 738--742.
\end{hangparas}

\begin{hangparas}{.25in}{1}
 Thibert-Plante, X., and S. Gavrilets. 2013. Evolution of mate choice and the so-called magic traits in ecological speciation. Ecol. Lett. 16(8): 1004--1013.
\end{hangparas}

\begin{hangparas}{.25in}{1}
Via, S.  2009. Natural selection in action during speciation. Proc. Natl. Acad. Sci. USA 106: 9939--9946.
\end{hangparas}

\begin{hangparas}{.25in}{1}
Wagner, G. P., and J. Zhang.  2011. The pleiotropic structure of the genotype-phenotype map: the evolvability of complex organisms. Nat. Rev. Genet. 12: 204--213.
\end{hangparas}

\begin{hangparas}{.25in}{1}
Yeaman, S. 2013. Genomic rearrangements and the evolution of clusters of locally adaptive loci. Proc. Natl. Acad. Sci. USA 110(19): E1743--E1751.
\end{hangparas}

\begin{hangparas}{.25in}{1}
Yeaman, S. 2015.  Local adaptation by alleles of small effect. Am. Nat. 186(Suppl 1): S74--S89.
\end{hangparas}

\begin{hangparas}{.25in}{1}
Yeaman, S., and S. P. Otto.  2011. Establishment and maintenance of adaptive genetic divergence under migration, selection, and drift. Evolution 65(7): 2123--2129.
\end{hangparas}

\begin{hangparas}{.25in}{1}
Yeaman, S. and M. C. Whitlock. 2011.  The genetic architecture of adaptation under migration-selection balance.  Evolution 65(7): 1897--1911.
\end{hangparas}

\newpage
\section{Tables}
\noindent {\bf Table I: }Parameters of the model, and the values used in our computer simulations. The symbol $^\star$ indicates the parameter values used for the results shown in the main text. Results for other parameter values are shown in {\bf Supplementary information}.

\begin{table*}[tbhp]
  \begin{tabular*}{\hsize}
 {@{\extracolsep{\fill}}lllllllll}
\multicolumn1l{Symbol}&\multicolumn1l{Explanation}&\multicolumn1l{Values}\cr
\hline
\multicolumn1l{$N$}&\multicolumn1l{Population size per patch}&\multicolumn1l{$1000^\star$, $200$}\cr
\multicolumn1l{$L$}&\multicolumn1l{Number of adaptive loci}&\multicolumn1l{$100^\star$, $2000$}\cr
\multicolumn1l{$m$}&\multicolumn1l{Migration rate}&\multicolumn1l{$0.1^\star$}\cr
\multicolumn1l{$\theta^{(k)}$}&\multicolumn1l{Optimal phenotype in population $k=1,2$}&\multicolumn1l{$\theta^{(1)}=2^\star,\quad\theta^{(2)}=-2^\star$}\cr
\multicolumn1l{$\sigma$}&\multicolumn1l{Selection parameter}&\multicolumn1l{$4^\star$, $3.5$, $2.5^\star$ }\cr
\multicolumn1l{$r$}&\multicolumn1l{Recombination rate}&\multicolumn1l{$0.0005^\star$, $0.001$}\cr
\multicolumn1l{$\mu$}&\multicolumn1l{Mutation rate}&\multicolumn1l{$2\cdot 10^{-5~\star}$, $10^{-4}$}\cr
\multicolumn1l{$\sigma_\mu$}&\multicolumn1l{Root mean square of mutation-effect sizes}&\multicolumn1l{$0.05^\star$, $0.05/\sqrt{20}$}\ \ \cr
\hline
\end{tabular*}
\end{table*}

\newpage
\section{Figure legends}

{\bf Figure 1.} Multi-locus model results. Panels {\bf A} and {\bf C}: temporal dynamics of extents of local genomic divergence (truncated to the range indicated by the  colour bar) in single stochastic realisations of the model for weak selection ({\bf A}) and for strong selection ({\bf C}). The grey lines show the corresponding total extents of genomic divergence (the values are given on the $y$ axis on the right). Panels {\bf B} and {\bf D}: correlations of extents of divergence at pairs of loci as a function of their distance (measured in units of the recombination rate $r$) averaged over $90$ independent realisations for the parameters in {\bf A} and {\bf C}, respectively. Correlations are colour coded (see the colour bar).  Grey lines show the corresponding total extents of divergence averaged over $90$ independent realisations. 
Other parameter values: selection parameter $\sigma=4$ (in {\bf A}, {\bf B}) or $\sigma=2.5$ (in {\bf C}, {\bf D}), population size $N=1000$, mutation rate $\mu=2\cdot10^{-5}$, root mean square of  mutation-effect sizes $\sigma_\mu=0.05$, migration rate $m=0.1$, recombination rate between a pair of adjacent loci $r=5\cdot 10^{-4}$, number of adaptive loci $L=100$. Note that the timescales in the upper panels differ from those in the bottom ones.

{\bf Figure 2.} Rate of gain and rate of loss in the two-locus gain-loss model with one weakly diverged locus ($D_{\rm w}=0.2$) and a more strongly diverged one ($D_{\rm s}=0.4$). Shown are the rates as a function of distance between the loci (measured in units of recombination rate $r$). Panels {\bf A} and {\bf B}: rate of gain at the weakly diverged locus (circles) and at the more strongly diverged locus (squares) for weak selection ({\bf A}), and for strong selection ({\bf B}). Dashed lines indicate the mutation rate $\mu$ (this rate corresponds to the rate at which a neutral mutation lands and fixates at a neutral locus in a diploid population of size $N$).  Panels {\bf C} and {\bf D}: corresponding rates of loss for the parameters in {\bf A} and {\bf B}, respectively. Circles and squares overlap in {\bf D}. 
Other parameter values: selection parameter $\sigma=4$ in {\bf A} and {\bf C} or $\sigma=2.5$ in {\bf B} and {\bf D}, population size in each deme $N=1000$, migration rate $m=0.1$, mutation-effect size $\epsilon=0.05$, recombination rate $r=5\cdot 10^{-4}$, mutation rate $\mu=2\cdot 10^{-5}$. Number of simulations used: $2\cdot10^6$ in {\bf A}, $5\cdot10^5$ in {\bf B}, $10^3$ in {\bf C}, and $200$ in {\bf D}.

{\bf Figure 3.} Rate of loss relative to rate of gain in the two-locus gain-loss model for weak selection  at different stages of divergence (different $D_{\rm s}$) as a function of distance between the loci (measured in units of the recombination rate $r$). Dash-dotted lines correspond to a ratio of unity. The arrow in each panel depicts an approximate location of the critical distance above which the rate of loss is larger than the rate of gain for the corresponding value of $D_{\rm s}$. Selection parameter: $\sigma=4$. Number of simulations used: $2\cdot10^6$ for the rate of gain, $1000$ for the rate of loss. Other parameter values are the same as in Fig.~2.

{\bf Figure 4.} Same as in Fig.~3, but for strong selection ($\sigma=2.5$). Both panels are for late stages of divergence ($D_{\rm s}=3.8$ in {\bf A}, and $D_{\rm s}=3.9$ in {\bf B}). Number of simulations used for the rate of gain: $5\cdot10^5$. Number of simulations used for the rate of loss: $200$ in {\bf A} and $1000$ in {\bf B}.  Other parameter values are the same as in Fig.~2. 
\newpage
\section{Figures}

\begin{figure}[tbhp]
\begin{center}
{\includegraphics[width=16.5cm,angle=0]{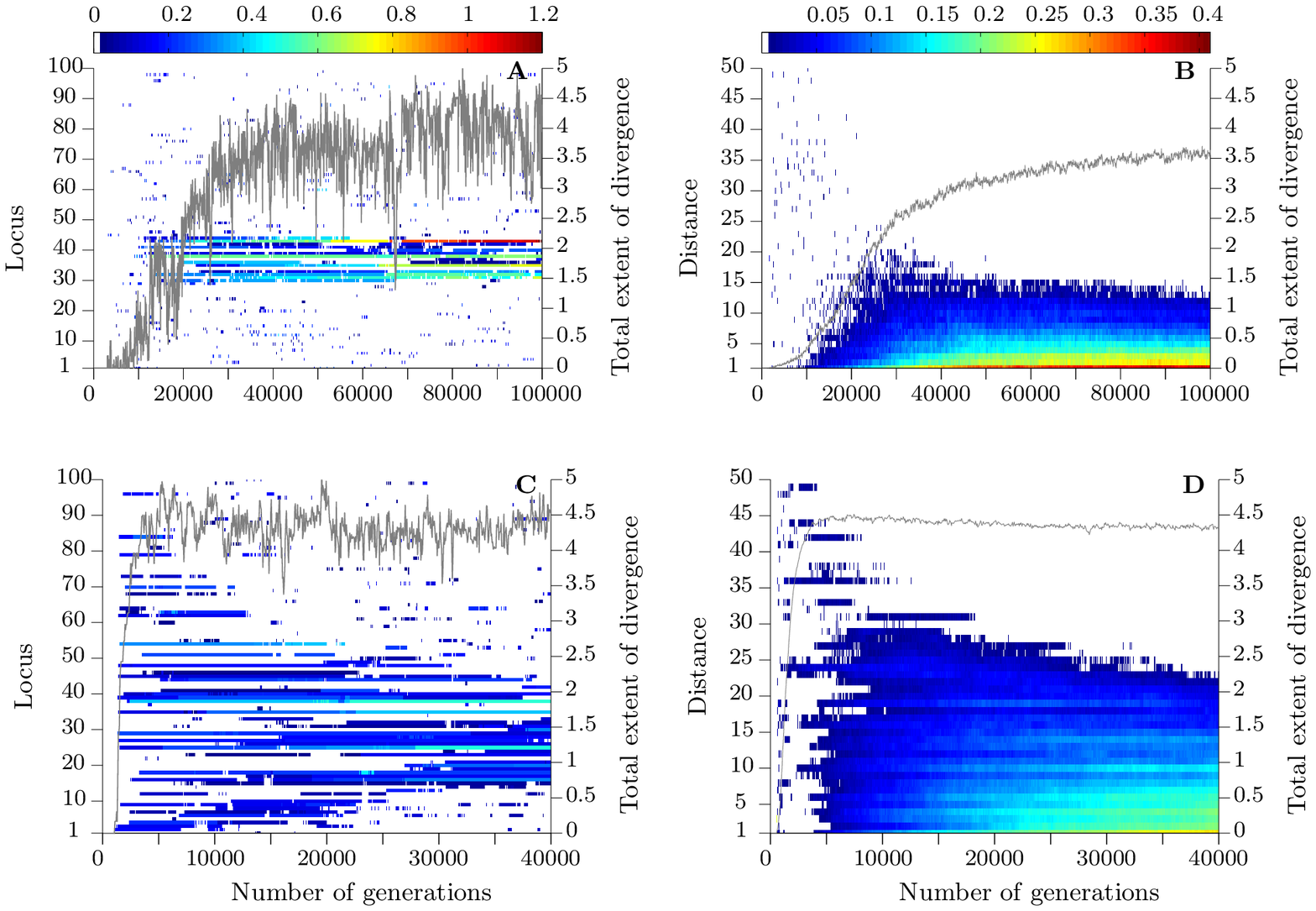}}
\caption{\label{fig:1}}
\end{center}
\end{figure}

\newpage

\begin{figure*}[tbhp]
\begin{center}
{\includegraphics[width=16.5cm,angle=0]{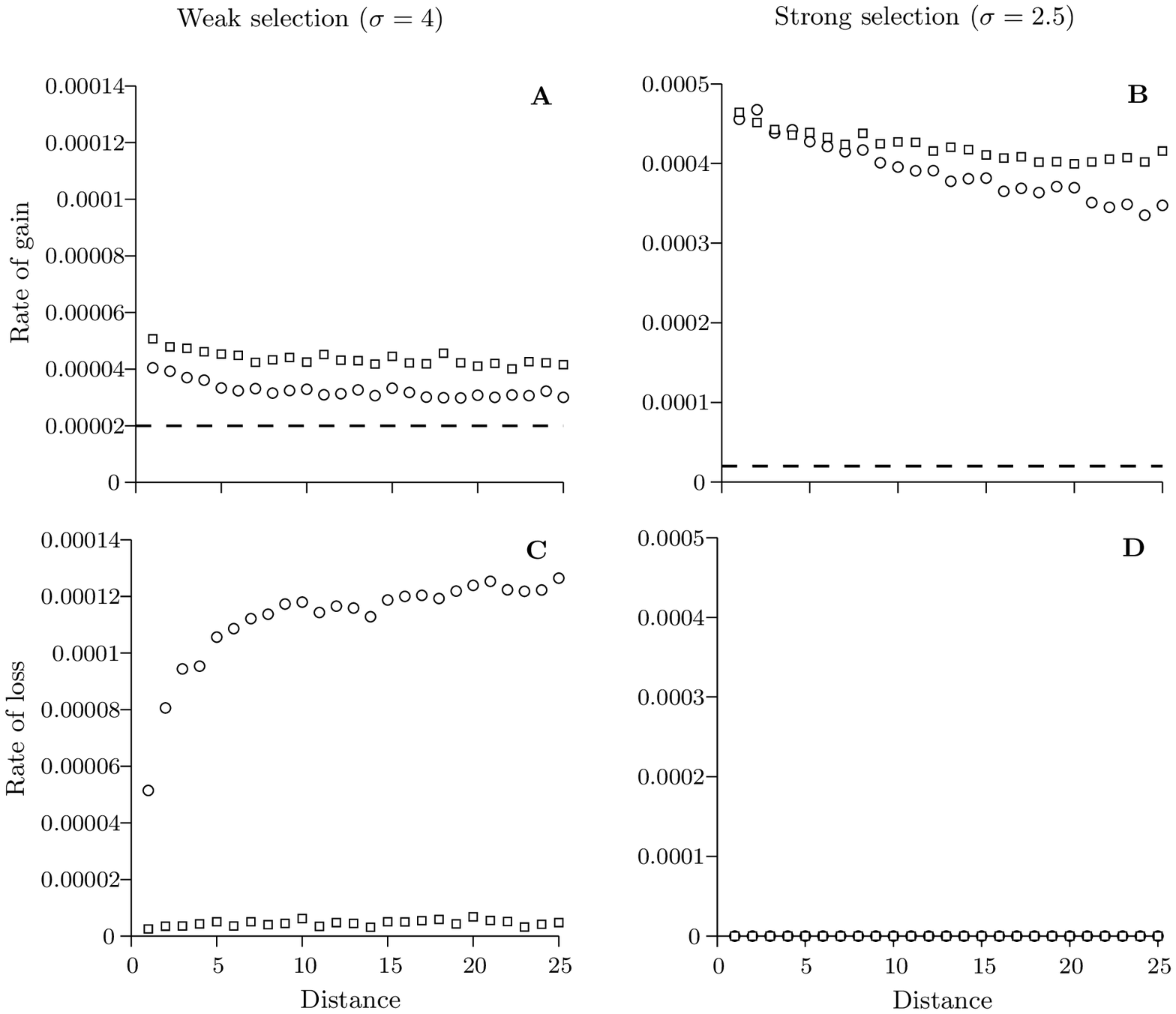}}
\caption{\label{fig:2}}
\end{center}
\end{figure*}

\newpage

\begin{figure}[tbhp]
\begin{center}
{ \includegraphics[width=7cm,angle=0]{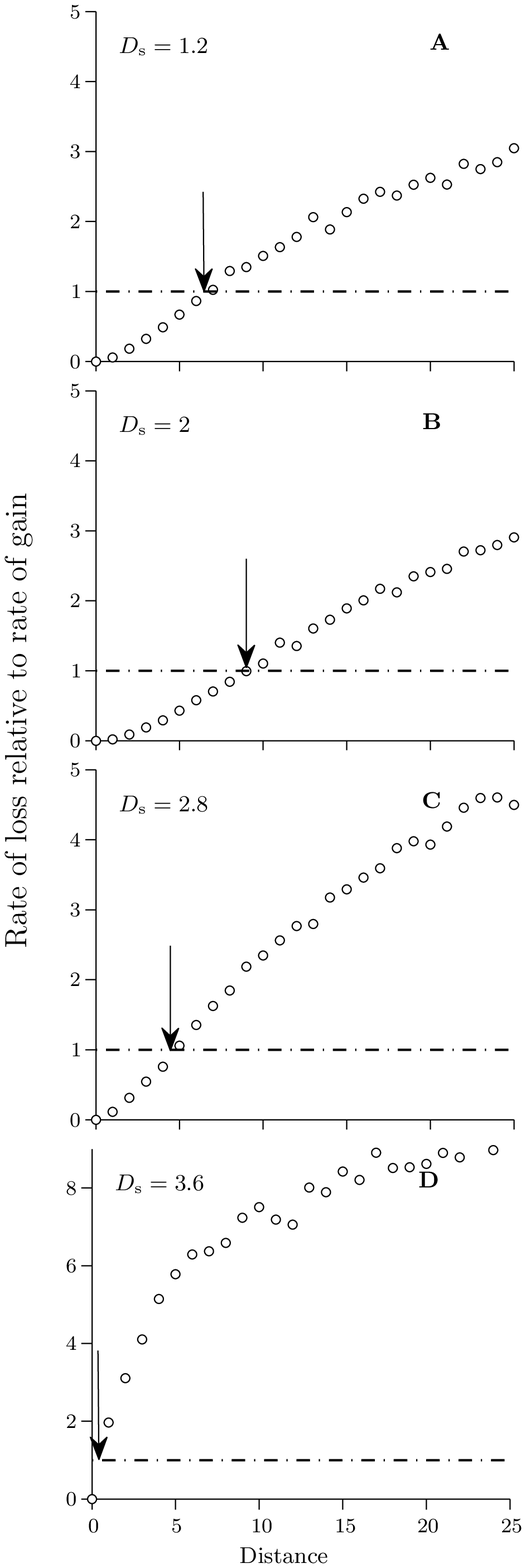}}
\caption{\label{fig:4} }
\end{center}
\end{figure}

\newpage

\begin{figure}[tbhp]
\begin{center}
{\includegraphics[width=7.5cm,angle=0]{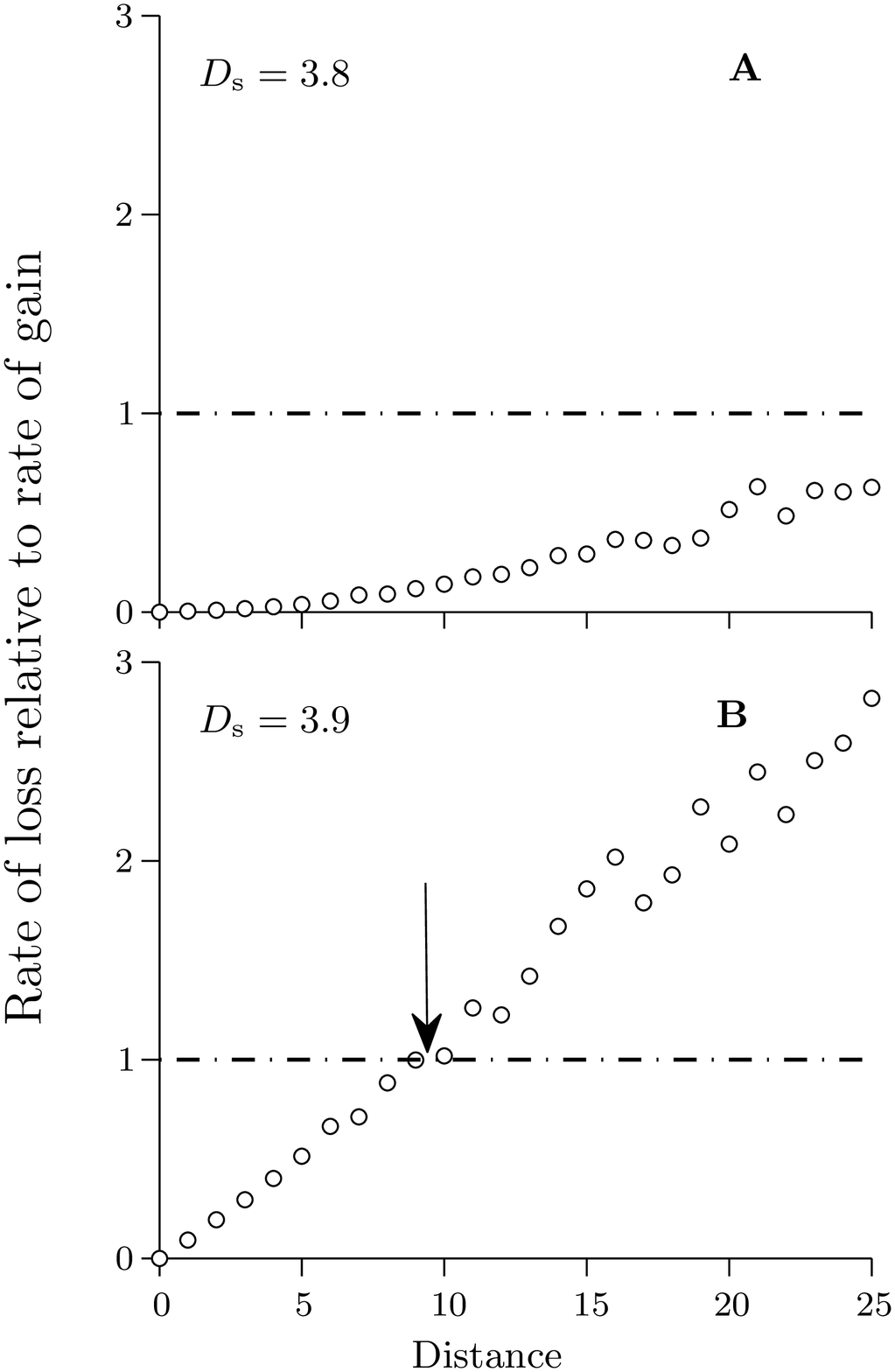}}
\caption{\label{fig:5} }
\end{center}
\end{figure}
\end{document}


\title{Supplementary information for ``A universal mechanism generating clusters of differentiated loci during divergence-with-migration"}
\author{Marina Rafajlovi{\'c}$^{1,2,*}$, Anna Emanuelsson$^{1}$, Kerstin Johannesson$^{3,2}$, Roger K. Butlin$^{4,2}$, and Bernhard Mehlig$^{1,2}$\\
$^1$\emph{\small Department of Physics, University of Gothenburg, SE-412 96 Gothenburg, Sweden}\\
$^2$\emph{\small The Linnaeus Centre for Marine Evolutionary Biology, University of Gothenburg, SE-405 30 Gothenburg, Sweden}\\
$^3$\emph{\small Department of Marine Sciences, University of Gothenburg, Tj\"arn\"o SE-452 96 Str\"omstad, Sweden}\\
$^4$\emph{\small Department of Animal and Plant Sciences, University of Sheffield, Sheffield S10 2TN, UK}\\
$^*$\emph{\small Corresponding author, {\tt marina.rafajlovic@physics.gu.se}}\\
}
\beginsupplement
\maketitle

\section{Details about selection parameters used in the main results}
In the main text, patterns of divergence for the model introduced in {\bf Materials and Methods} are shown for two values of the selection parameter $\sigma$, that is, for $\sigma=4$, and $\sigma=2.5$. Selection is weaker in the former than in the latter case. In either case, the optimal trait values $\theta^{(1)}$ and $\theta^{(2)}$ in the two demes are set to $\theta^{(1)}=-\theta^{(2)}=2$. Therefore, for $\sigma=4$ we find that a perfectly adapted individual in one deme  
experiences a fitness disadvantage of about $0.39$ in comparison to perfectly adapted individuals  in the other deme, and the fitness disadvantage of the first generation hybrids between perfectly adapted individuals in the two demes is about $0.12$ (also in relation to locally perfectly adapted individuals). Note that in earlier stages of divergence, the fitness disadvantage of the first generation hybrids in comparison to locally favourable individuals (that have not reached their optima) in either deme is smaller than $0.12$. In this respect, the extent of divergent selection in all stages of divergence under $\sigma=4$ in our model corresponds to weak levels of selection under the model of Feder et al. (2012) ($s_{\rm o}<0.12$ in their model). Under the stronger selection ($\sigma=2.5$) in our model, the corresponding fitness disadvantage of individuals that are perfectly adapted for the opposite deme, and of the first generation hybrids  are about $0.72$, and $0.27$, respectively. 

 We also note that when selection is weak, that is, when $\sigma^2$ is large in comparison to the distance between the optima $\theta^{(1)}-\theta^{(2)}$, our fitness function (Eq. (1) in the main text) reduces to that used by Yeaman and Whitlock (2011), with $\gamma=2$, $\theta^{(1)}=-\theta^{(2)}=\theta$, and $\Phi=2\theta^2/\sigma^2$ in their model. In particular, the selection strength corresponding to setting $\sigma=2.5$ in our model is  similar to that used by Yeaman and Whitlock (2011) for their parameter $\Phi=0.75$ (see above). 

Finally, the standard deviation $\sigma_\mu$ of the Gaussian distribution from which mutation-effect sizes are drawn is set to $\sigma_\mu=0.05$ in the main text. With this choice, we find that in the first population (with the positive optimal phenotype) the selective advantage of a heterozygote with allele-effect sizes $0|0.05$ (the latter corresponding roughly to one mutation of effect-size $0.05$ landing on an allele of effect size $0$) over the homozygote with allele-effect sizes $0|0$  is about $0.006$ for the weaker selection tested ($\sigma=4$), whereas for the stronger selection ($\sigma=2.5$) it is about $0.016$ (which is approximately three times larger than under the weaker selection).

\section{Two-locus establishment model}

We use a two-locus establishment model  to analyse the importance of the establishment advantage of mutations landing close to a diverged locus in comparison to those landing at a distance. In these simulations we assume that one of the two loci has diverged before a new mutation lands in the genome. This locus is assumed to have alleles of effect sizes $Y_{\rm s} >0$ and $-Y_{\rm s} $ that are in a migration-selection balance. The initial frequencies of these two allele-effect sizes in the two populations are determined using a set of recursive deterministic equations (see {\bf S4}). Note that in populations of infinite size, a balance between allele-effect sizes $Y_{\rm s} $ and $-Y_{\rm s} $ (both having nonzero frequencies, see {\bf S4}) will be established and maintained under any migration rate due to the symmetries  assumed in the model (Yeaman and Otto 2011) (but the allele frequencies depend on the migration rate). With these settings, the extent of local genomic divergence $D_{\rm s} $ at this locus prior to mutation is $D_{\rm s} =4Y_{\rm s} $.
In addition, the second locus is assumed to have alleles of effect size zero prior to mutation. Therefore the extent of local genomic divergence $D_{\rm w} $ at this locus is equal to zero. Thereafter we assume that a mutation lands on the undifferentiated locus (or the diverged one, see below) in the population where it is locally beneficial, and we simulate the dynamics of genotype frequencies under drift to estimate the probability that the mutation successfully establishes in the populations. In these simulations, the mutation-effect size $\epsilon>0$ is assumed to be fixed and equal to the standard deviation $\sigma_\mu$ of the mutation-effect size distribution. Because the mutation of size $\epsilon>0$ is beneficial in the first population (with the positive optimal phenotype) we assume it lands in the first population. Thereafter, new mutations are not allowed. We  neglect mutations landing in the population where they are locally deleterious because these are much less likely to establish successfully in comparison to mutations landing in the population where they are locally favourable. Each simulation is advanced until either the mutant allele experiences extinction, or 
until it becomes most common (frequency $>50\%$) at the locus  in the population where it is beneficial. In the latter case, a successful establishment event is noted. We run at least $10^5$ such independent simulations, and  the establishment probability is estimated as the proportion of successful establishment events among all independent runs made. We estimate the establishment probabilities of a new mutation landing at various recombination distances $r_j$ from the diverged locus. The values of $r_j$ are chosen as $r_j=j r$ ($j=0,\ldots,50$), where $r=0.0005$ corresponds to the recombination distance between adjacent loci set in Fig.~1. When $j=0$, a mutation lands on the diverged locus (`stacking' \`a la Yeaman and Whitlock (2011)). In this case, we additionally assume that a mutation lands on a locally favourable allele, giving rise to the mutant allele that is locally advantageous in comparison to either allele at this locus prior to the mutation. When $j=50$, the distance between the two loci corresponds to a half of the total recombination distance assumed in Fig. 1. Varying $Y_{\rm s} $ from zero to unity we approximate different stages of divergence from no divergence to perfect local adaptation in both populations. Results obtained in this model are shown in Figs.~S2-S3.

\section{Two-locus gain-loss model}

In this appendix we explain the assumptions used in the two-locus gain-loss model that is introduced in the main text. In this model the two loci are assumed to be at a recombination distance $r_j>0$. Both loci are assumed to be differentiated initially, one with a stronger and the other with a weaker extent of divergence. Each locus has two alleles with effect sizes that are symmetric around zero ($Y_{\rm s} >0$  and $-Y_{\rm s} $ at the more strongly diverged locus, and $0<Y_{\rm w} <Y_{\rm s} $ and $-Y_{\rm w} $ at the weakly diverged one).  We set the allele-effect sizes at the weakly diverged locus  to $Y_{\rm w}=\sigma_\mu$, and $-Y_{\rm w}$. We choose this value as a representative beneficial mutation-effect size in the first population in a situation when mutation-effect sizes are drawn from a Gaussian distribution with a zero mean and a standard deviation $\sigma_\mu$ (assuming that when divergence starts all loci have alleles with effect sizes zero). Mutation-effect sizes much smaller than this value appear with a higher probability, but they suffer from a lower establishment probability. By contrast, mutation-effect sizes larger than this value appear with a  much smaller probability. The initial haplotype frequencies are assumed to be equal to those in the deterministically expected stable steady state of the system (see {\bf S4}). After initialisation we run two sets of simulations. In one we aim to estimate the rates of local loss at the two loci (neglecting new mutations). In the other we aim at estimating the rate of local gain upon introducing a mutation. These two sets of simulations are described in the main text, where we also show and discuss the results obtained under the gain-loss model.

\section{Deterministic approximation for a two-locus model}\label{app:twoloci}
In this appendix we list  a set of recursive two-locus deterministic equations for adaptive divergence that we use to determine haplotype frequencies at the start of simulations of the establishment, and gain-loss models introduced in {\bf Materials and Methods} in the main text. The deterministic approximation for the dynamics of haplotype frequencies is valid in the limit of infinitely large populations. 

The main assumptions of the model of adaptive divergence are introduced in the main text. The populations are assumed to be diploid and of equal size $N$ that is constant over time. For purposes of this appendix, we assume here that $N\rightarrow\infty$. The environmental conditions are assumed to be different in the two demes, so that a given phenotype is under divergent selection. The optimal phenotype in the first (second) deme is denoted by $\theta^{(1)}$ ($\theta^{(2)}$), and we assume that $\theta^{(1)}>0$, and $\theta^{(2)}=-\theta^{(1)}$. In the two-locus model, the phenotype of an individual is assumed to be determined by the diploid genotype at  two adaptive loci. Each allele at a given locus is assigned an allele-effect size by which it additively contributes to the phenotype. Selection is assumed to be soft, so that a contribution of  individual $i$ with phenotype $z$ to the gamete pool in deme $k$ is proportional to the fitness $w_i^{(k)}$ of this individual  relative to the fitness of all individuals in this deme, where

\begin{equation}\label{eq:fitness}
w_i^{(k)}={\rm e}^{-\frac{(z-\theta^{(k)})^2}{2\sigma^2}}\,\,.
\end{equation}

\noindent The strength of selection is determined by the parameter $\sigma$ in such a manner that selection is weaker when $\sigma$ is larger, and vice versa. In the model, individuals firstly migrate to the opposite deme at a rate $m$ per individual, generation. Thereafter, random mating, recombination and selection occur  locally within each deme. Recombination is assumed to occur at a rate $r$ per gamete, individual, generation. When $r=0$, the model described corresponds to a single-locus model.  

 In this appendix we assume that each locus has two possible alleles, and we aim at estimating the haplotype frequencies in the steady state of the system. The effect sizes of these alleles are assumed to be symmetric around zero, and we denote them by $Y_{\rm s} >0$ and $-Y_{\rm s} $ at one locus, and $Y_{\rm w} >0$ and $-Y_{\rm w} $ at the other locus. We assume that $Y_{\rm s} $, and $Y_{\rm w} $ are advantageous over $-Y_{\rm s} $, and $-Y_{\rm w} $, respectively in the first population. The opposite is true in the second population. When $Y_{\rm s} =Y_{\rm w} $ the two loci do not differ in the extents of their divergence, whereas for $Y_{\rm s} >Y_{\rm w} $, the first locus has a higher extent of divergence  than the second one. In what follows, we use a deterministic approximation to find the haplotype (and allele) frequencies at the two loci in the stable steady state.

When two divergent populations are initialised with allele-effect sizes $x_1=Y_{\rm s} $ and $x_2=-Y_{\rm s} $ at one locus, and with $y_1=Y_{\rm w} $ and $y_2=-Y_{\rm w} $ at the other locus, a deterministic approximation shows that each locus establishes a stable dimorphism (see also Yeaman and Otto (2011)). This conclusion can be arrived at by iterating a system of recursive equations for the evolution of frequencies $p^{(k)}_{x_i,y_j;\tau}$, of haplotypes $x_i,y_j$ ($i,j=1,2$)  in the two populations ($k=1,2$) from generation $\tau$ to generation $\tau+1$. The dynamics are fully determined by a set of six equations. For simplicity, however, we show here the corresponding equation for $p^{(1)}_{x_1,y_1;\tau+1}$ noting that the remaining five equations are obtained similarly:

\begin{align}
p^{(1)}_{x_1,y_1;\tau+1}=&\Bigl[(1-m) \Bigl(p^{(1)}_{x_1,y_1;\tau}\Bigr)^2+m \Bigl(p^{(2)}_{x_1,y_1;\tau}\Bigr)^2\Bigr]\frac{w^{(1)}_{x_1,y_1|x_1,y_1}}{\langle w^{(1)}_\tau \rangle}\nonumber\\
&+\Bigl[(1-m) p^{(1)}_{x_1,y_1;\tau} p^{(1)}_{x_1,y_2;\tau} +m p^{(2)}_{x_1,y_1;\tau}p^{(2)}_{x_1,y_2;\tau} \Bigr]\frac{w^{(1)}_{x_1,y_1|x_1,y_2}}{\langle w^{(1)}_\tau \rangle}\nonumber\\
&+\Bigl[(1-m) p^{(1)}_{x_1,y_1;\tau} p^{(1)}_{x_2,y_1;\tau} +m p^{(2)}_{x_1,y_1;\tau}p^{(2)}_{x_2,y_1;\tau} \Bigr]\frac{w^{(1)}_{x_1,y_1|x_2,y_1;\tau}}{\langle w^{(1)}_\tau \rangle}\nonumber\\
&+r\Bigl[(1-m)  p^{(1)}_{x_1,y_2;\tau} p^{(1)}_{x_2,y_1;\tau} +m p^{(2)}_{x_1,y_2;\tau}p^{(2)}_{x_2,y_1;\tau} \Bigr]\frac{w^{(1)}_{x_1,y_2|x_2,y_1}}{\langle w^{(1)}_\tau \rangle}\nonumber\\
&+(1-r)\Bigl[(1-m) p^{(1)}_{x_1,y_1;\tau} p^{(1)}_{x_2,y_2;\tau} +m p^{(2)}_{x_1,y_1;\tau}p^{(2)}_{x_2,y_2;\tau} \Bigr]\frac{w^{(1)}_{x_1,y_1|x_2,y_2}}{\langle w^{(1)}_\tau \rangle}\,\,.\label{eq:1}
\end{align}  

\noindent Here, $\langle w^{(1)}_\tau \rangle$ denotes the average fitness  of parents in the first population in generation $\tau$ and it is given by 

 \begin{equation}
\langle w^{(1)}_\tau\rangle=\sum_{i=1}^2\sum_{j=1}^2\sum_{l=1}^2\sum_{a=1}^2 \Bigl[(1-m) p^{(1)}_{x_i,y_j;\tau}p^{(1)}_{x_l,y_a;\tau}+m p^{(2)}_{x_i,y_j;\tau}p^{(2)}_{x_l,y_a;\tau}\Bigr]w^{(1)}_{x_i,y_j|x_l,y_a}\,\,,
\end{equation}

\noindent where the subscripts $i$ and $j$ denote alleles at the two loci at one chromosome (similarly, $l=1,2$ and $a=1,2$ are used for the corresponding pair at the other chromosome). The superscript $k=1,2$ stands for the first and second population, respectively. The fitnesses $w^{(k)}_{x_i,y_j|x_l,y_a}$   ($i,j,k,l,a=1,2$) are  given by Eq.~(\ref{eq:fitness}) with $z=x_i+y_j+x_l+y_a$.

Using Eq.~(\ref{eq:1}) and the remaining five equations of the system, we find recursively the state at which the system eventually relaxes within a predetermined numerical precision. As a stopping condition for finding this state we require that during $1000$ successive generations neither of the allelic frequencies change by more than $10^{-8}$. The maximum number of generations for finding the steady state is set to $10^5$. For all parameter values tested, we find a stable dimorphism with nonzero allele frequencies at both loci as well as linkage disequilibrium between loci, the extent of which can be determined according to the haplotype frequencies obtained. 

In a special case when $r=0$, and the system is initialised with haplotypes $Y_{\rm s}, Y_{\rm w} $ and $-Y_{\rm s},-Y_{\rm w} $ (effectively corresponding to allele-effect sizes $Y_{\rm s} +Y_{\rm w} $ and $-Y_{\rm s} -Y_{\rm w} $ at a single locus), a deterministic approximation shows that, independently of the migration rate, the stable steady state of the system corresponds to both alleles having nonzero frequencies (but the actual frequencies depend on the migration rate and the selective advantage of locally beneficial allele). This stems from the analysis performed in Yeaman and Otto (2011) upon assuming in their model symmetric migration, and that the alleles labelled by $A$, and $a$ have effect sizes $-Y_{\rm s} -Y_{\rm w} $, and $Y_{\rm s} +Y_{\rm w} $, respectively. A stable dimorphism in the single-locus case is a consequence of the symmetries assumed in the model. As mentioned already,  allele frequencies in the stable state attained depend on the migration rate, so that the frequency of locally favourable alleles is smaller under stronger migration.

We finalise this appendix by noting  that we use the equations given above only to estimate the initial haplotype (and hence allele) frequencies in simulations  under the two-locus establishment model, and under the two-locus gain-loss model introduced in the main text. All simulations are otherwise stochastic and performed under random genetic drift.

\section{References}
\begin{hangparas}{.25in}{1}
Feder, J. L., R. Gejji, S. Yeaman, and P. Nosil. 2012. Establishment of new mutations under divergence and genome hitchhiking. Philos. Trans. R. Soc. Lond. B Biol. Sci. 367(1587): 461--474.
\end{hangparas}

\begin{hangparas}{.25in}{1}
Yeaman, S., and S. P. Otto.  2011. Establishment and maintenance of adaptive genetic divergence under migration, selection, and drift. Evolution 65(7): 2123--2129.
\end{hangparas}

\begin{hangparas}{.25in}{1}
Yeaman, S. and M. C. Whitlock. 2011.  The genetic architecture of adaptation under migration-selection balance.  Evolution 65(7): 1897--1911.
\end{hangparas}

\newpage
\section{Supplementary Legends}

{\bf Figure S1.} Same as in Fig.~1{\bf A} in the main text, but here patterns from two different realisations are shown.  For clarity, grey lines here depict the total extent of divergence in intervals of $250$ generations (whereas Fig.~1{\bf A} shows all measures in intervals of $50$ generations). For the explanation and parameter values used, refer to Fig.~1{\bf A} in the main text.

{\bf Figure S2.} Results of the two-locus establishment model. Shown is the probability of establishment of a new mutation of a fixed size landing at an undifferentiated locus as a function of the distance between this locus and the locus that is differentiated prior to the mutation (measured in units of the recombination rate $r$). The establishment probability at distance zero corresponds to the mutation landing at the already differentiated locus. Dashed lines show the probability $1/(2N)$ of fixation of a neutral mutation at a neutral locus in a diploid population of size $N$. 
 Panels differ by the extent of divergence $D_{\rm s} $ at the more strongly diverged locus prior to the mutation.  Selection is weaker in {\bf A}, {\bf C}, {\bf E}, {\bf G} ($\sigma=4$) than in {\bf B}, {\bf D}, {\bf F}, {\bf H} ($\sigma=2.5$). 
Remaining parameter values: population size in each deme $N=1000$, migration rate $m=0.1$, recombination rate $r=5\cdot 10^{-4}$, mutation-effect size $\epsilon=0.05$, $10^6$ independent simulations in {\bf A}, {\bf C}, {\bf E}, {\bf G}, and $10^5$ in {\bf B}, {\bf D}, {\bf F}, {\bf H}.

{\bf Figure S3.} Establishment bias in the two-locus establishment model. Shown is the integral of the establishment probability over distances outside of a given region around the diverged locus relative to the integral over distances within the region, as a function of the proportion that the region accounts for (out of $L=100$ loci). Dashed line indicates the ratio of unity. Note that in {\bf A} the ratio is, as expected, below unity (approximately $0.95$) when the proportion of the region around the diverged locus is equal to $0.5$, but this is difficult to observe due to the scale of the $y$-axis used. Selection is weaker in {\bf A} ($\sigma=4$) than in {\bf B} ($\sigma=2.5$). 
Parameters: mutation-effect size $\epsilon=0.05$, the extent of divergence at the already diverged locus $D_{\rm s}=0.4$, population size in each deme $N=1000$, migration rate $m=0.1$, $10^6$ independent simulations in {\bf A}, and $10^5$ in {\bf B}.

{\bf Figure S4.} Patterns of divergence under the parameter values similar to those in Fig.~1{\bf A}, {\bf B} in the main text, but here the recombination distance between adjacent loci is two times larger ($r=0.001$). Shown are the results from a single realisation in {\bf A}, and averages over $54$ independent realisations in {\bf B}. Remaining parameter values are the same as in Fig.~1{\bf A} in the main text.

{\bf Figure S5.} Effect of drift. Same as in Fig.~1 in the main text, but here we contrast the results obtained under a small population size ($N=200$, {\bf A}, {\bf B}), and a large population size ($N=1000$, {\bf C}, {\bf D}).  The mutation rate is set so that its value scaled by the corresponding population size is equal in the two cases ($\mu=10^{-4}$ in {\bf A} and {\bf B}, or $\mu=2\cdot10^{-5}$ in {\bf C} and {\bf D}).  In both cases, the selection parameter is  $\sigma=3.5$. Other parameter values are the same as in Fig.~1 in the main text.

{\bf Figure S6.} Same as in Fig.~3 in the main text but for the selection parameter $\sigma=3.5$ (corresponding to that used in Fig.~S5).   For the explanation of the figure refer to  Fig.~3 in the main text. Panels differ by the initial extent of divergence $D_{\rm s}$ at the more strongly diverged locus. Population size: $N=200$ (in {\bf A}, {\bf C}, {\bf E}, {\bf G} and {\bf I}), and $N=1000$ (in {\bf B}, {\bf D}, {\bf F}, {\bf H} and {\bf J}).  Mutation rate: $\mu=10^{-4}$ (in {\bf A}, {\bf C}, {\bf E}, {\bf G} and {\bf I}), and $2\cdot 10^{-5}$ (in {\bf B}, {\bf D}, {\bf F}, {\bf H} and {\bf J}). Mutation-effect size: $\epsilon=0.05$. For each parameter combination, the rate of gain is estimated based on $10^5$ independent simulations. The rate of loss is estimated using $10^3$ independent simulations ({in \bf A}, {\bf C}, {\bf E}, {\bf G}, {\bf I}), or $500$ simulations (in {\bf B}, {\bf D}, {\bf F}, {\bf H}, {\bf J}). Other parameters are the same as in Fig.~S5.

{\bf Figure S7.} Patterns of divergence under the parameter values similar to those in Fig.~1{\bf C} in the main text but with $20$ times more adaptive loci ($L=2000$), and $20$ times smaller variance $\sigma_\mu^2$ of mutation-effect sizes. Panel {\bf A}: the extent of divergence at all loci in a single stochastic realisation of the model. The solid line  shows the total extent of divergence in the underlying single realisation of the model (the values are depicted on the $y$-axis on the right). Panel {\bf B}: correlations at pairs of loci as a function of time and the distance between them (measured in units of recombination rate $r$) averaged over $10$ independent realisations. The corresponding average total extent of divergence is shown by the solid line. Panel {\bf C}: same as in {\bf A}, but  depicting a cluster of diverged loci that accounts for most of the total extent of divergence in the realisation shown. 
Loci are assumed to reside on two chromosomes, so that the loci labelled $1,\ldots,1000$ are on one chromosome, and loci labelled $1001,\ldots,2000$ are on the other.  Root mean square of mutation-effect sizes: $\sigma_\mu=0.05/\sqrt{20}$. 
Remaining parameters are the same as in Fig.~1{\bf C}.

{\bf Figure S8.} Same as in Fig.~3 in the main text but for the parameters corresponding to those in Fig.~S7. Panels differ by the initial extent of divergence $D_{\rm s}$ at the more strongly diverged locus. Mutation-effect size: $\epsilon=0.05/\sqrt{20}$. The extent of divergence  $D_{\rm w} $ at the weakly diverged locus is set to $D_{\rm w}=4\epsilon$ (i. e. $D_{\rm w}=0.2/\sqrt{20}$). The rate of gain is estimated using $10^5$ independent simulations. The rate of loss is based on $200$ simulations.
Other parameters are the same as in Fig.~S7.

{\bf Figure S9.} Patterns of divergence for the parameter values corresponding to those in Figure 1 in the main text, but here mutation-effect sizes are drawn from an exponential distribution mirrored around zero (that is, positive and negative effects are assumed to be equally likely).  For the explanation of the results shown refer to the caption of Figure 1 in the main text. Number of independent realisations in panels {\bf B}, {\bf D}: $50$. Remaining parameter values are the same as in Fig. 1 in the main text.

{\bf Figure S10.} A comparison between patterns of divergence in a single stochastic realisation of the model, but shown using two different measures for the extent of divergence at locus $j$, that is, in {\bf A} we use the measure $D_l$ introduced in the main text (the total extent of divergence is equal to $\sum_{l=1}^L D_l$), and in {\bf B} we use instead twice the difference between average allele-effect sizes at locus $l$ in the two populations (the average extent of divergence is equal to the sum of average allele-effect sizes at all $L$ loci simulated). All parameter values correspond to those in Figure 1{\bf C} in the main text, but here the result of a different stochastic realisation is shown. For further explanation of the results shown refer to the caption of Figure 1{\bf C} in the main text.

\newpage

\section{Supplementary Figures}

\begin{figure}[tbhp]
\centerline{
{\includegraphics[width=16.5cm,angle=0]{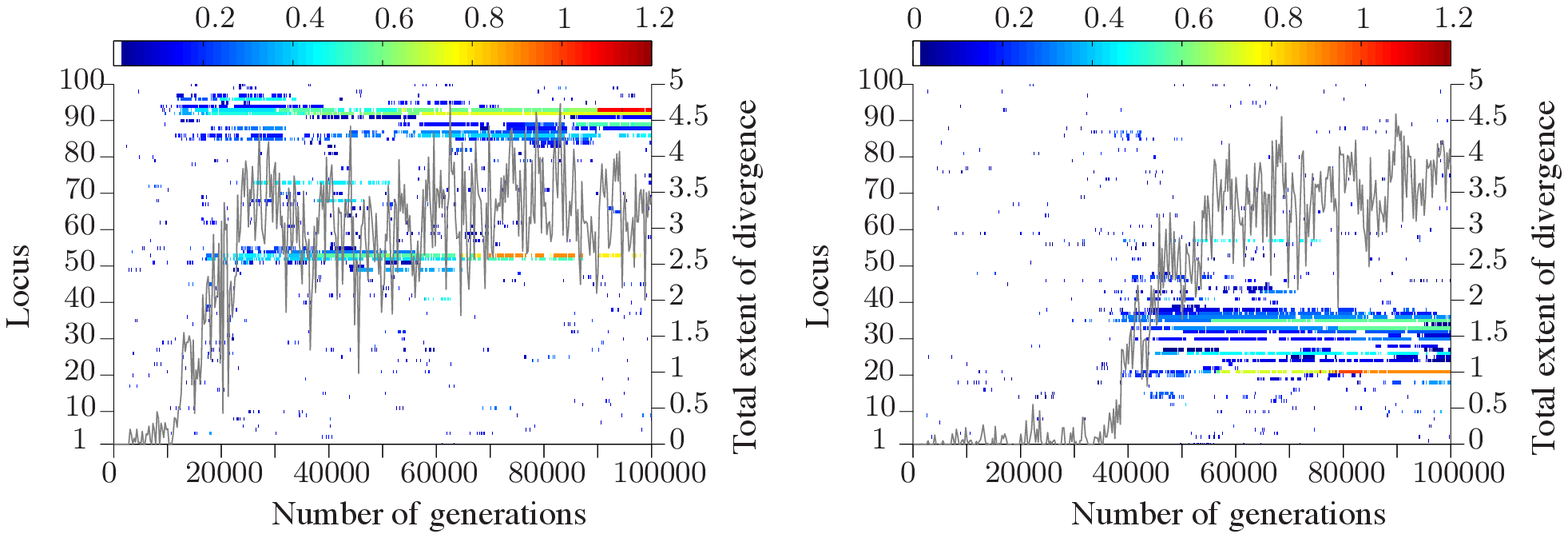}}
}
\caption{\label{fig:difr_real_s4p0} }
\end{figure}

\newpage
\begin{figure}[tbhp]
\centerline{
{\includegraphics[width=16.5cm,angle=0]{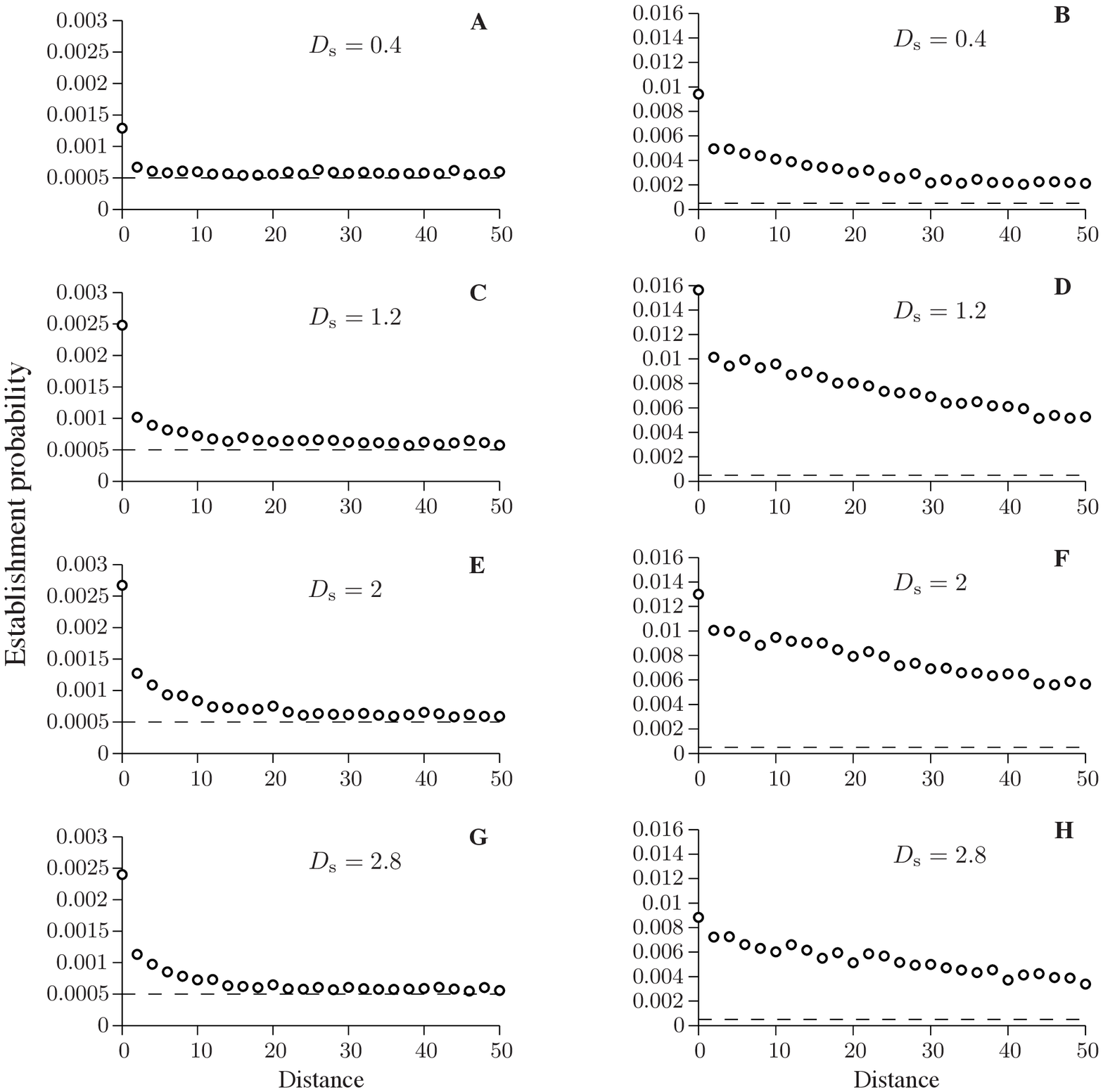}}}
\caption{\label{fig:IN_est} }
\end{figure}

\newpage
\begin{figure}[tbhp]
\centerline{
{\includegraphics[width=16.5cm,angle=0]{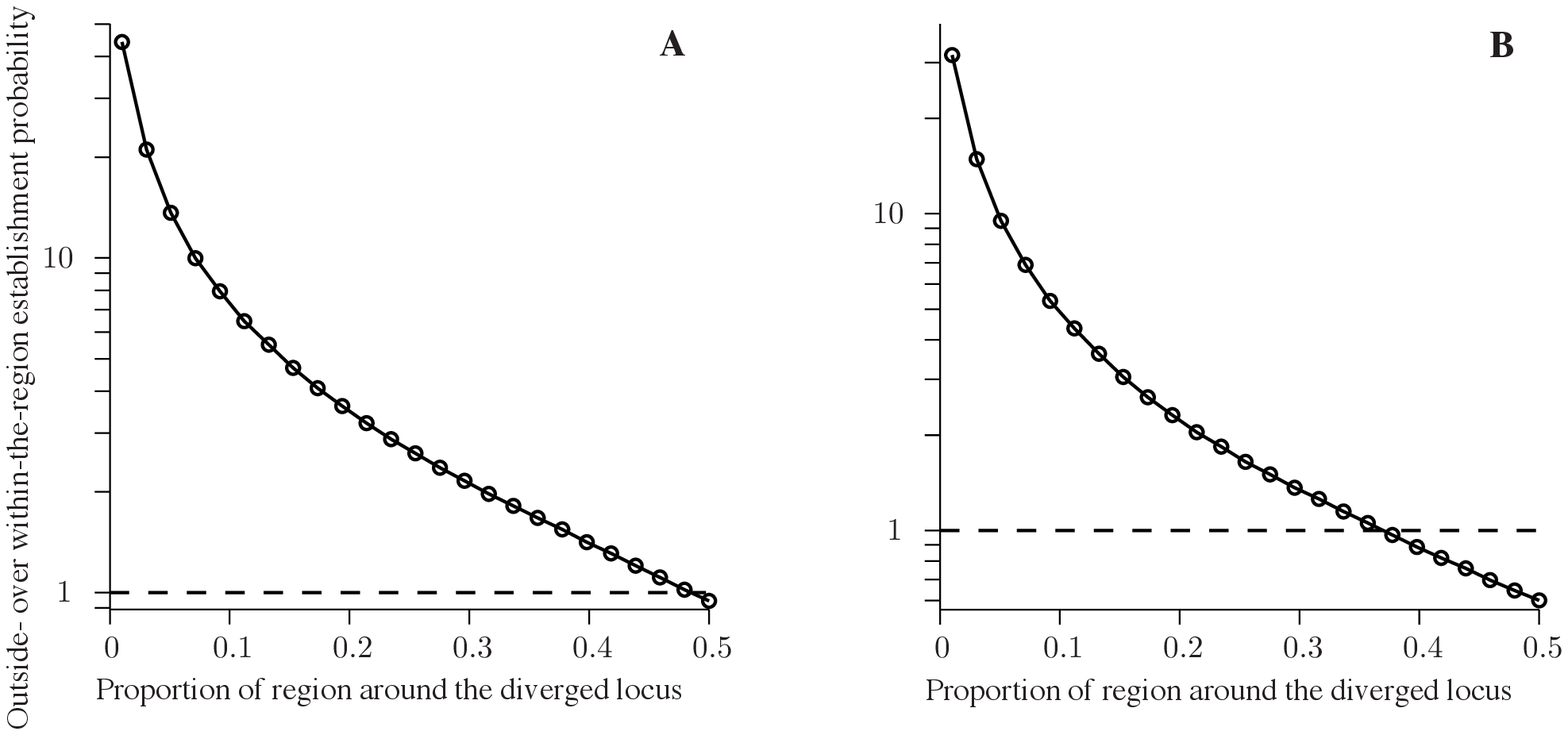}}
}
\caption{\label{fig:IN_est1}}
\end{figure}

\newpage
\begin{figure}[tbhp]
\centerline{
{\includegraphics[width=16.5cm,angle=0]{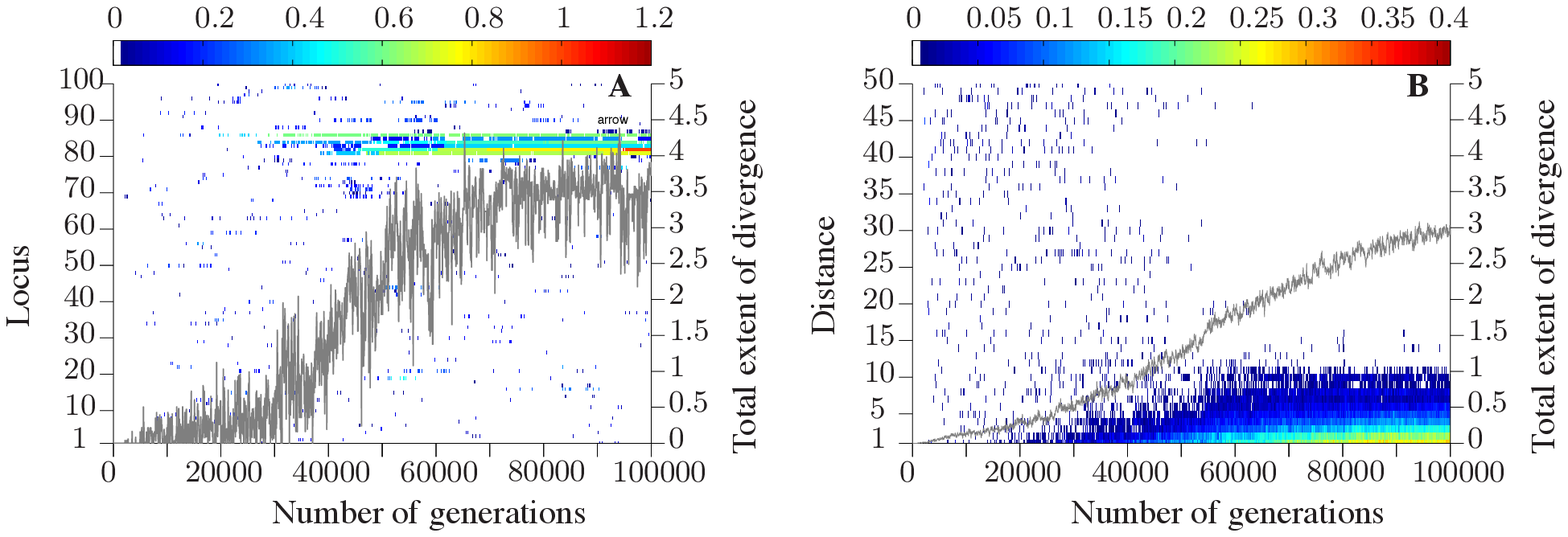}}
}
\caption{\label{fig:r0p001}}
\end{figure}

\newpage
\begin{figure}[tbhp]
\centerline{
{\includegraphics[width=16.5cm,angle=0]{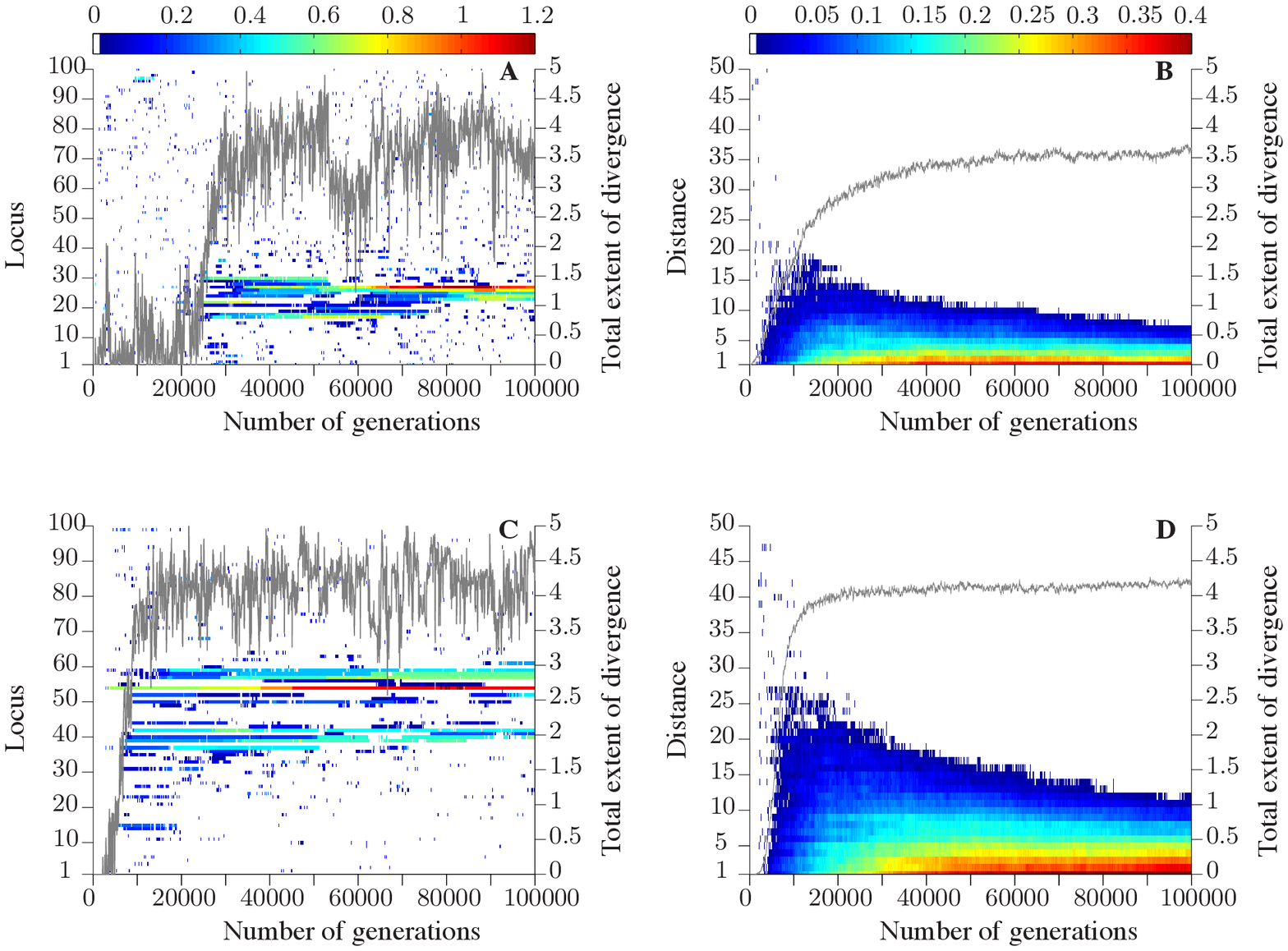}}
}
\caption{\label{fig:drift}}
\end{figure}

\newpage
\begin{figure}[tbhp]
\begin{center}
 {\includegraphics[width=16.5cm,angle=0]{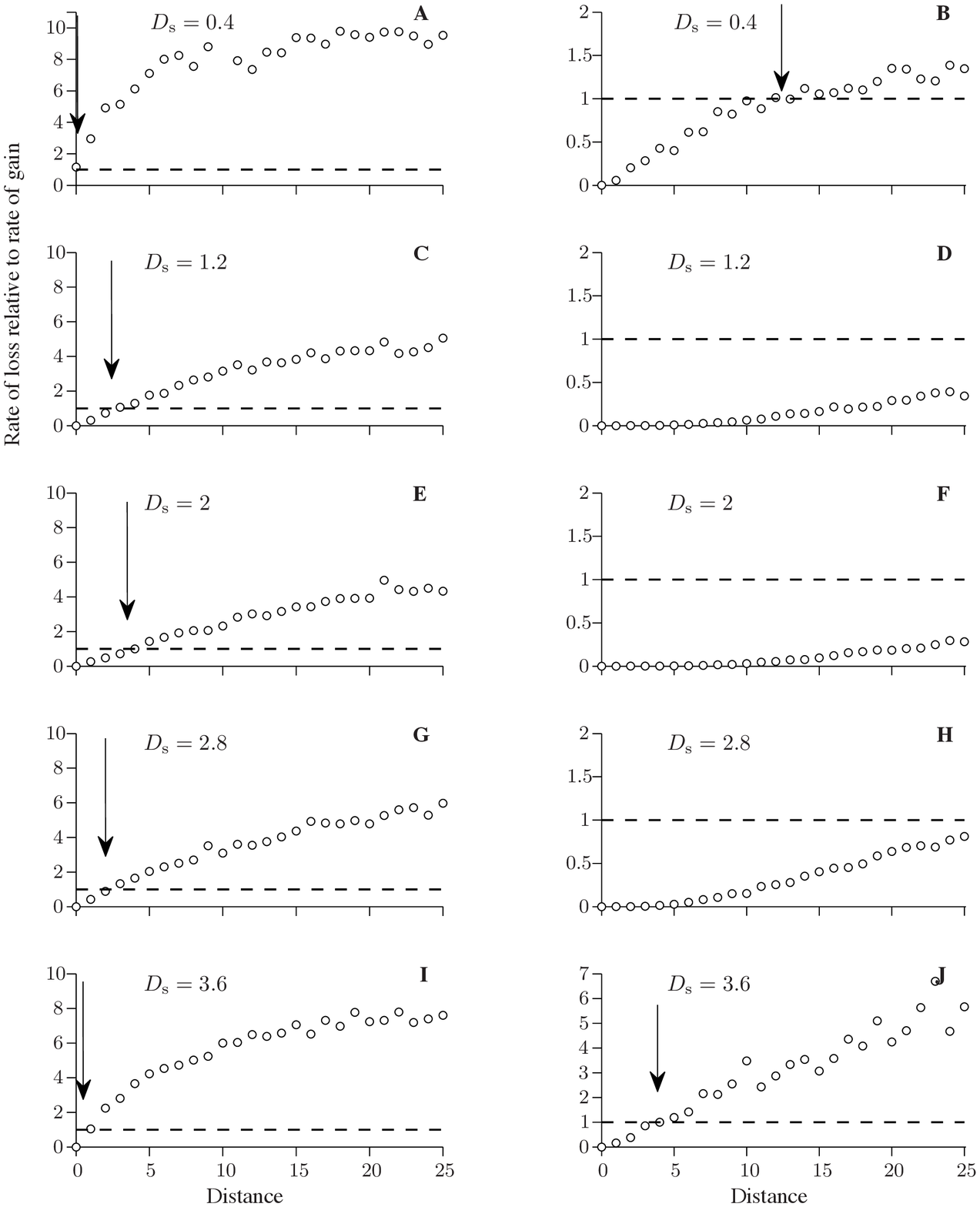}}
\caption{\label{fig:loss_gain}}
\end{center}
\end{figure}

\newpage
\begin{figure}[tbhp]
\centerline{
{\includegraphics[width=16.5cm,angle=0]{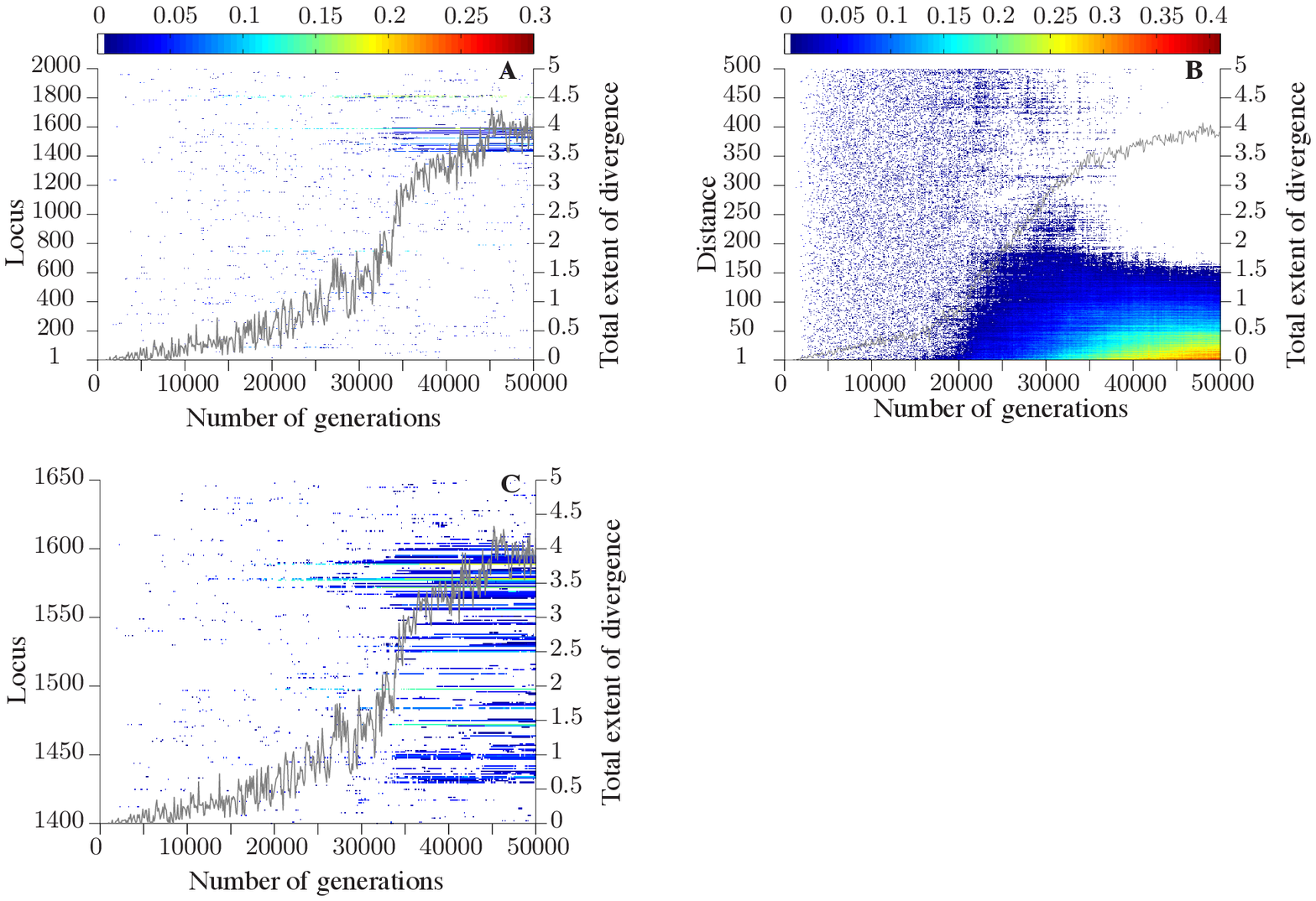}}
}
\caption{\label{fig:2000Loci_scMu}}
\end{figure}

\newpage
\begin{figure}[tbhp]
\centerline{
{\includegraphics[width=8cm,angle=0]{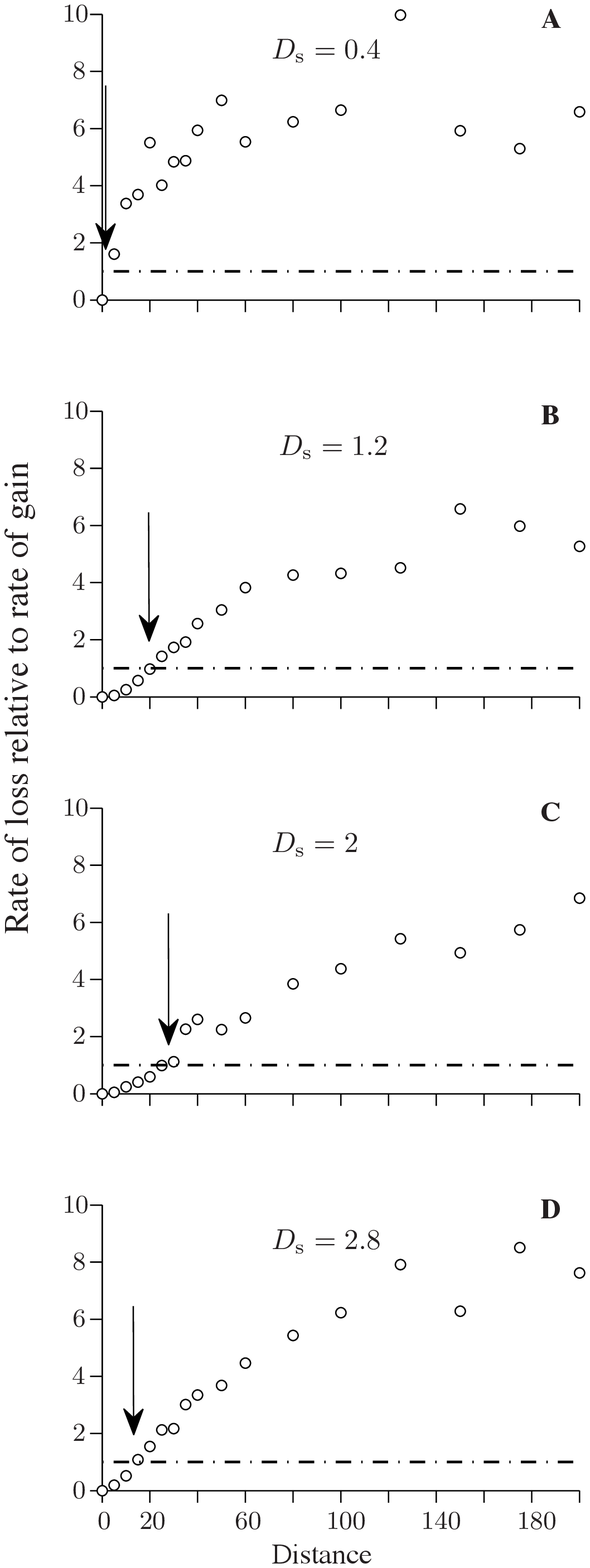}}
}
\caption{\label{fig:loss_gain_L2000} }
\end{figure}

\protect{\newpage
\begin{figure}[tbhp]
\centerline{
{\includegraphics[width=16cm,angle=0]{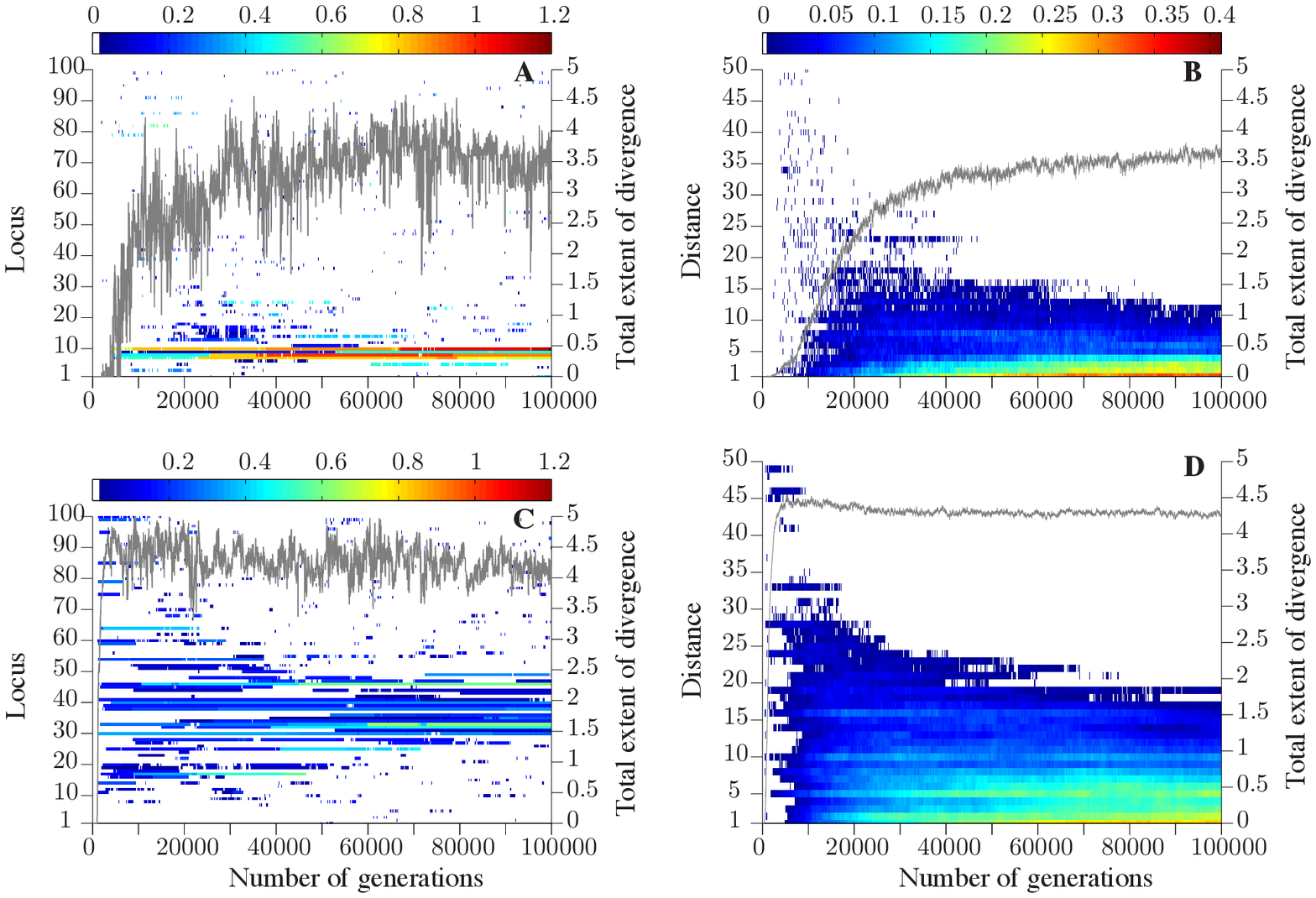}}
}
\caption{\label{fig:ExpMut4} }
\vspace*{10cm}
\end{figure}
}

\protect{\newpage
\begin{figure}[tbhp]
\centerline{
{\includegraphics[width=16cm,angle=0]{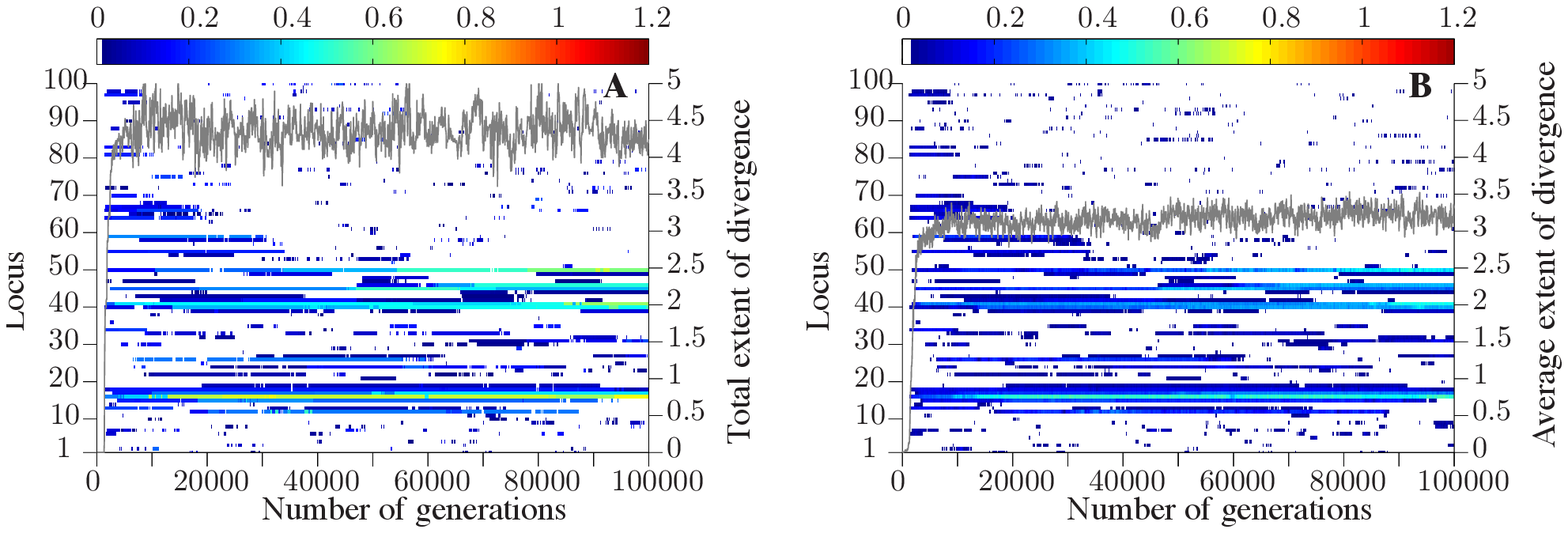}}
}
\caption{\label{fig:avAls2p5} }
\end{figure}
}